\documentclass[11pt,sort&compress]{elsarticle}
\usepackage[paper=a4paper,marginratio={1:1,5:6}]{geometry}
\makeatletter
\def\ps@pprintTitle{%
 \let\@oddhead\@empty
 \let\@evenhead\@empty
 \def\@oddfoot{\centerline{\thepage}}%
 \let\@evenfoot\@oddfoot}
\makeatother
\usepackage{natbib}
\usepackage[hyperref]{xcolor}
\definecolor{darkgreen}{rgb}{0.01, 0.75, 0.24}
\definecolor{darkblue}{HTML}{2B66D3}
\usepackage[colorlinks=true,
            linkcolor=darkblue,
            urlcolor=darkblue,
            citecolor=darkblue,linkcolor=darkblue]{hyperref}

\usepackage{amsmath}
\usepackage{amsfonts}
\usepackage{eqnarray}
\usepackage{siunitx}
\usepackage{slashed}
\usepackage{accents}
\usepackage{amssymb}
\usepackage{mathrsfs}
\usepackage{mathtools}
\usepackage[utf8]{inputenc}
\usepackage[T1]{fontenc}
\let\oldbibliography\thebibliography
\renewcommand{\thebibliography}[1]{%
  \oldbibliography{#1}%
  \setlength{\itemsep}{1.4pt}%
}

\usepackage[cal=boondox,calscaled=1]{mathalfa}
\DeclareMathAlphabet{\bbvar}{U}{BOONDOX-ds}{m}{n}
\DeclareMathAlphabet{\mathsl}{\encodingdefault}{\familydefault}{m}{sl}

\usepackage[defaultsans,scale=0.9]{opensans}
\usepackage{psfrag}
\usepackage{pstool}
\usepackage{caption}
\usepackage{tabularx}
\usepackage{multirow}
\usepackage{pbox}

\usepackage{graphicx}
\usepackage{subcaption}
\usepackage{relsize}

\newcommand{\qq}[1]{``#1''} 

\newcommand{\di}{\mathrm{d}}
\usepackage{tensor}
\newcommand{\ou}[3]{\tensor{#1}{^{#2}_{#3}}}
\newcommand{\uo}[3]{\tensor{#1}{_{#2}^{#3}}}

\newcommand{\vo}[3]{\tensor[^{\mathnormal{#1}}]{#2}{^{#3}}}
\newcommand{\vu}[3]{\tensor[^{\mathnormal{#1}}]{#2}{_{#3}}}
\newcommand{\vou}[4]{\tensor[^{\mathnormal{#1}}]{#2}{^{#3}_{#4}}}

\newcommand{\ubar}[1]{\underaccent{\bar}{#1}}

\newcommand{\I}{\mathrm{i}} 
\newcommand{\E}{\mathrm{e}} 
\newcommand{\CC}{\mathrm{cc.}} 
\newcommand{\C}{\mathbb{C}}

\newcommand{\R}{\mathbb{R}}

\newcommand{\eref}[1]{(\ref{#1})}

\DeclareMathAlphabet{\bbgreek}{U}{bbold}{m}{n}

\newcommand{\mtext}[1]{\text{\it #1}}
\usepackage{tikz-cd}
\usetikzlibrary{arrows.meta, automata,
                positioning,
                quotes}
\usepackage[low-sup]{subdepth}

\newcommand\vpm{\mathbin{\vcenter{\hbox{
  \oalign{\hfil$\scriptstyle+$\hfil\cr
          \noalign{\kern-.3ex}
          $\scriptscriptstyle({-})$\cr}}}}}
\DeclareMathAlphabet{\sfit}{OT1}{\sfdefault}{m}{it}
\DeclareMathAlphabet{\sfbfit}{OT1}{\sfdefault}{sb}{it}
\DeclareMathAlphabet{\mathsf}{OT1}{\sfdefault}{sb}{n}

\definecolor{darkgreen}{rgb}{0.01, 0.75, 0.24}
\definecolor{darkblue}{HTML}{2B66D3}

\usepackage[multiple, flushmargin]{footmisc}
\let\originalleft\left
\let\originalright\right
\renewcommand{\left}{\mathopen{}\mathclose\bgroup\originalleft}
\renewcommand{\right}{\aftergroup\egroup\originalright}

\usepackage{scalerel}

\newcommand{\dbarvar}{{\mathrm{d}\mkern-7.5mu\lower.18ex\hbox{$\textasciitilde$}\mkern-1.5mu}}

\renewcommand{\emph}[1]{{\it #1}}

\begin{document}

\begin{abstract}
We consider the phase space of the Maxwell field as a simplified framework to study the quantisation of holonomies (Wilson line operators) on lightlike (null) surfaces. Our results are markedly different from the spacelike case. On a spacelike surface, electric and magnetic fluxes each form a commuting subalgebra. This implies that the holonomies commute. On a lightlike hypersurfaces, this is no longer true. Electric and magnetic fluxes are no longer independent.  To compute the Poisson brackets explicitly, we choose a regularisation. Each path is smeared into a thin ribbon. In the resulting holonomy algebra, Wilson lines commute unless they intersect the same light ray. . 
We compute the structure constants of the holonomy algebra and show that they depend on the geometry of the intersection and the conformal class of the metric at the null surface. Finally, we propose a quantisation. 
The resulting Hilbert space shows a number of unexpected features. First, the holonomies become anti-commuting Grassmann numbers.
Second, for pairs of Wilson lines, the commutation relations can continuously interpolate between fermionic and bosonic relations. Third, there is no unique ground state. The ground state depends on a choice of framing of the underlying paths. 
\end{abstract}%
\title{Fermionisation of the Aharonov--Bohm Phase on the Lightfront}
\author{Carolina Sole {Panella} and Wolfgang {Wieland}}
\address{Institute for Quantum Gravity, Theoretical Physics III, Department of Physics\\Friedrich-Alexander-Universität Erlangen-Nürnberg\\Staudstraße 7, 91052 Erlangen, Germany
\\{\vspace{0.5em}\normalfont 24 November 2025}}

\maketitle
\vspace{-1.5em}
\hypersetup{
  linkcolor=black,
  urlcolor=black,
  citecolor=black
}
{\tableofcontents}\vspace{-0.5em}
\hypersetup{
  linkcolor=darkblue,
  urlcolor=darkblue,
  citecolor=darkblue,
}
\begin{center}{\noindent\rule{\linewidth}{0.4pt}}\end{center}\newpage
\section{Introduction}

\noindent There are a number of  results that have improved the case for a quasi-local quantisation of gravitational null initial data.  
These developments are scattered among different communities, connecting earlier research on the 
loop representation of quantum geometry \cite{status,ashtekar,rovelli,thiemann,zakolec,Gambini_Pullin_1996}, spinfoams \cite{alexreview,Rovelli:2014ssa,Kaminski:2009fm,Dittrich:2018xuk,Dittrich:2014ala,Asante:2020qpa}, 
 non-expanding and isolated horizons \cite{Ashtekar:aa,Ashtekar:2004aa,Ashtekar:2001is,Lewandowski:2002ua,Lewandowski:1999zs,Ashtekar:2021wld,Ashtekar:2024stm,Ashtekar:2021kqj}, quantum reference frames \cite{Rovelli:1990pi,Giacomini:2019aa,Hardy:2019cef,Vanrietvelde:2018pgb,Hoehn:2019fsy,delaHamette:2020dyi,Loveridge2018} to edge modes \cite{Balachandran:1994up,Carlip:1996yb,Freidel:2015gpa,Wieland:2017zkf,Wieland:2017cmf,Wieland:2021vef,Speranza:2017gxd,Freidel:2021cjp,Freidel:2021fxf,Freidel:2020xyx,Carrozza:2021gju,Kabel:2023jve,Freidel:2023bnj,Giesel:2024xtb,Fewster:2024pur,Fewster:2025ijg}, subsystem observables \cite{Giddings:2019hjc,Donnelly:2016auv,Donnelly:2016rvo,Donnelly:2015hta,Donnelly:2017jcd} and local and asymptotic symmetries of the gravitational field 
\cite{AshtekarNullInfinity,Barnich:2011mi,Strominger:2017zoo,Pasterski:2015zua,Pasterski:2021raf,Wieland:2020gno,Freidel:2021dfs,Ruzziconi:2025fuy}.\smallskip

In this paper, we study a simplified version of the programme. We consider the phase space of a local subsystem of the Maxwell field on a lightlike (null) hypersurface. This surface is embedded into a four-dimensional background spacetime. The background metric is fixed and no interactions between the Maxwell field and gravity are considered. We introduce the classical Poisson algebra of observables for such a subsystem and compute the algebra between $U(1)$ Wilson line operators, which are the exponentials of the Aharonov--Bohm phases. The resulting commutation relations are markedly different from the spacelike case. Wilson lines no longer commute. They commute unless they intersect the same light ray. The commutation relations are determined by structure constants that depend on the conformal class of the metric at the null surface. This is different from what happens in a topological field theory, where the structure constants are independent of the metric (up to a framing ambiguity) \cite{Wittenreal,ChernSimonsBook}. The difference occurs because Maxwell's theory is not topological, but couples to the conformal class of the metric. Finally, we propose a lattice-type quantisation, in which we observe a fermionisation of the $U(1)$ holonomies. Each Wilson line behaves as an anti-commuting Grassmann number on the null surface.\smallskip

The paper is organised as follows. \hyperref[sec2]{Section 2} is a basic review in which we discuss three topics: the covariant phase space approach \cite{AshtekarNullInfinity,1987CrnkovicWitten,Ashtekar:1987hia,Wald:1999wa}, the Poisson algebra between electric fluxes and $U(1)$ holonomies (compactified magnetic fluxes) on a spatial Cauchy surface and the corresponding lattice quantisation in terms of charge network states \cite{Ashtekar:1991my,Ashtekar:1991mz,Varadarajan:1999it}.
In \hyperref[sec3]{Section 3}, we introduce the classical phase space of a local subsystem of the Maxwell field on a null initial surface. This requires some basic elements of the geometry of null surfaces; we provide a brief summary of  the Newman--Penrose formalism \cite{newmanpenrose,penroserindler}, before then computing the symplectic structure. The phase space splits into radiative modes in the interior and edge modes  at the boundaries of the subsystem domain \cite{Donnelly:2016auv,Freidel:2023bnj}. The main results are developed in \hyperref[sec4]{Section 4} and \hyperref[sec5]{Section 5}, where we compute the classical Poisson algebra between Wilson lines on null surfaces and propose a quantisation of the resulting holonomy algebra. Finally, there is a brief summary and an \hyperref[appdx]{Appendix}, where we present some supplementary material on the mode expansion of the phase space observables for the Maxwell field on a local lightfront.

\section[Basic review: $\boldsymbol{U(1)}$ holonomy-flux algebra and charge network states]{Basic review: $U(1)$ holonomy-flux algebra and charge network states}\label{sec2}
\noindent In this section, we review some introductory material about the Hilbert space of a $U(1)$ lattice gauge theory on spatial hypersurfaces \cite{PhysRevD.11.395,PhysRevD.19.619,Varadarajan:1999it,Ashtekar:1991my,Ashtekar:1991mz,Bojowald:1999fw,Drobinski:2017kfm,Magnifico:2020bqt}, which provides a simplified test bed for the spin network representation of quantum geometry \cite{Smolin:1992wj,Varadarajan:2018tei,Zarate:2025erv,Bakhoda:2024mth,Bakhoda:2020ril,Bakhoda:2020fiy,Thiemann:2022all,Sahlmann:2024pba}. Our main purpose here is to recall some basic elements of the resulting charge network representation. The quantisation on null initial surfaces will be markedly different (see \hyperref[sec4]{Section 4} and \hyperref[sec5]{Section 5}).

\paragraph{Units and conventions} Our notation follows the conventions of \cite{wald,ashtekar}. The action for the free electromagnetic field in a curved background spacetime $(\mathcal{M},g_{ab})$ is built from the four-dimensional Faraday tensor $\vu{4}{F}{ab}=2\nabla_{[a}{}\vu{4}{A}{b]}$ for the vector potential $\vu{4}{A}{a}$, which serves as the configuration variable, and the background metric $g_{ab}$, i.e.\
\begin{equation}
S[\vu{4}{A}{}]=-\frac{1}{4e^2}\int_{\mathcal{M}}d^4v_g \vu{4}{F}{ab}\vo{4}{F}{ab}=-\frac{1}{2e^2}\int_{\mathcal{M}}\star \vu{4}{F}{}\wedge \vu{4}{F}{},
\end{equation}
where $d^4v_g$ is the volume element (a four-form) for the metric and $\star \vu{4}{F}{}$ is the Hodge dual of the Faraday tensor.\footnote{Our conventions are $[d^4v_g]_{abcd}=\varepsilon_{abcd}$, where $\varepsilon_{abcd}$ is the Levi-Civita tensor, and $[\star {}^4F]_{ab}=\tfrac{1}{2}\uo{\varepsilon}{ab}{cd}{}^4F_{cd}$.  The metric signature is $(-$$+$$+$$+)$, $\nabla_a$ is the covariant derivative and $a,b,c,\dots$ are abstract indices.} Units are chosen such that the e.g.\ $U(1)$ covariant derivative of a charged scalar field $\phi$ is given by $\mathcal{D}_a\phi:=\nabla_a\phi-\I\, \vu{4}{A}{a}\phi$. Thus, $e^2$ has dimensions $[e^2]=[\hbar^{-1}]$ (the action has dimensions of $\hbar$) and the coupling to an external current $j^a$ is given by adding the term $e^{-1}j^a\vu{4}{A}{a}$ to the Lagrangian.

\paragraph{Phase space} The variation of the action for fixed background metric ($\delta g_{ab}=0$) determines the equations of motion in the bulk and the pre-symplectic potential at the boundary,
\begin{align}
\delta S[\vu{4}{A}{}]=\frac{1}{e^2}\int_{\mathcal{M}}(\di\star \vu{4}{F}{})\wedge \delta \vu{4}{A}{}-\frac{1}{e^2}\int_{\partial\mathcal{M}}\star \vu{4}{F}{}\wedge\delta \vu{4}{A}{}.
\end{align}
If we impose the field equations, the first term vanishes and the second term provides the definition of the canonical momenta through the Hamilton--Jacobi functional. This is to say that the boundary term defines the symplectic potential, which is the Cauchy surface integral of the symplectic current
\begin{equation}
\theta=-e^{-2}\star \vu{4}{F}{}\wedge\bbvar{d}\vu{4}{A}{},\label{sympl-curr}
\end{equation}
where $\bbvar{d}$ denotes the exterior derivative on field space.\footnote{In here, the field space exterior derivative commutes with the exterior derivative on spacetime, i.e.\ $\di\bbvar{d}=\bbvar{d}\di$. Other authors prefer to combine $\di$ and $\bbvar{d}$ into an extended exterior derivative $\mathbf{d}=\di+\bbvar{d}$, in which case we would obtain a relative minus sign $\di\bbvar{d}=-\bbvar{d}\di$ to maintain $\mathbf{d}^2=0$, see e.g.\ \cite{Riello:2022din,Blohmann:2022yqo}. }  Consider then a spacelike Cauchy surface $\Sigma$. We integrate the symplectic current along $\Sigma$ and obtain the (pre-)symplectic potential on the covariant phase space \cite{AshtekarNullInfinity,1987CrnkovicWitten,Ashtekar:1987hia,Wald:1999wa},
\begin{equation}
\Theta_\Sigma=\int_\Sigma\theta=-e^{-2}\int_\Sigma\tilde{E}^a\bbvar{d}{A}_a.
\end{equation}
In here, we introduced the pair of canonical variables
\begin{align}
A_a&=\vu{4}{A}{b}\uo{h}{a}{b}\big|_\Sigma,\\
\tilde{E}^{a}&=d^3v_h\,\vou{4}{F}{a}{b}n^b\big|_\Sigma\equiv d^3v_h\,E^a,
\end{align}
where $\uo{h}{a}{b}=\delta^b_a+n_an^b$ is the projector onto $\Sigma$, $n^a$ denotes the future pointing surface normal to $\Sigma$ and $d^3v_h$ is the volume element with respect to the three metric $h_{ab}=g_{ab}+n_an_b$. The Poisson brackets on the resulting kinematical phase space are
\begin{equation}
\big\{\tilde{E}^a(x),A_b(y)\big\}_\Sigma=-e^2\ou{h}{a}{b}\tilde{\delta}_\Sigma(x,y).
\end{equation}
All other brackets among the canonical variables vanish. The physical phase space is obtained by imposing the Gauss constraint
\begin{equation}
D_a{E}^a=0,\label{Gsslaw}
\end{equation}
where $D_a$ is the covariant derivative intrinsic to $\Sigma$.


\paragraph{Holonomy-flux variables} The kinematical phase space of field configurations $(A_a,E^a)$ on $\Sigma$ admits a natural discretisation in terms of holonomy-flux variables \cite{PhysRevD.11.395,PhysRevD.19.619,Bojowald:1999fw,Drobinski:2017kfm,Magnifico:2020bqt}. Consider  a triangulation\footnote{In what follows, the restriction to a triangulation is not necessary. The generalisation to arbitrary graphs---ordered lists of paths, not necessarily dual to triangulations---was developed in \cite{Ashtekar:1993wf,Ashtekar:1995zh}.} $\Delta$ of $\Sigma$ consisting of tetrahedra $(T_1,T_2,\dots)$, triangles $(t_1,t_2,\dots)$, edges $(e_1,e_2,\dots)$ and vertices $(v_1,v_2,\dots)$ and the dual cellular complex $\Delta^\ast$, which consists of nodes $n_1,n_2,\dots$ (dual to tetrahedra), links $(l_1,l_2,\dots)$ (dual to triangles), faces $(f_1,f_2,\dots)$ (dual to edges) and bubbles $(b_1,b_2,\dots)$ (dual to vertices). To each link $l$ in the triangulation, we assign a holonomy (i.e.\ the exponential of $\I$\,$\times$\,the Aharonov--Bohm phase)
\begin{equation}
h_l[A]=\E^{-\I\int_l A}.
\end{equation}
In addition, we also introduce an electric flux $\mathcal{E}^l$ for each link $l$, which is the integral of the electric field over the triangle $t(l)$ dual to $l$,\footnote{Links and triangles are oriented: if $(X^a,Y^a)$ is a positively oriented dyad on the triangle $t(l)$ dual to the link $l$ and $Z^a$ is a tangent vector to $l$ that aligns with the orientation of $l$, the triple $(X^a,Y^a,Z^a)$ is positively oriented on $\Sigma$.}
\begin{equation}
\mathcal{E}^l[E]=\int_{t(l)}E,\label{fluxdef}
\end{equation}
where $E$ is the two-form, which is the dual of the electric field: $E_{ab}=\varepsilon_{abc}E^c$, where $\varepsilon_{abc}=n^f\varepsilon_{fabc}$ is the Levi-Civita tensor on $\Sigma$.\smallskip

Given the fundamental Poisson brackets, it is now immediate to show that the smeared variables close under the Poisson brackets,
\begin{equation}
\big\{\mathcal{E}^l,h_{l'}\big\}_\Sigma=-\I\,e^{2}\,\varepsilon(l,{l'})\,h_{l'},
\end{equation}
where $\varepsilon(l,l')=1$ if $l=l'$, $\varepsilon(l,l')=-1$ if $l^{-1}=l'$ and zero in all other cases. All other Poisson brackets among the $U(1)$ holonimies and electric fluxes vanish.\smallskip

\paragraph{Charge networks} 
For simplicity, we consider here the case in which the number of links in the triangulation is finite, i.e.\ $L_\Delta=\{l_1,\dots, l_N\}$ is the set of links, $N< \infty$.
 The resulting phase space $T^\ast U(1)^N$ of electric fluxes $\{\mathcal{E}^l\}_{l\in L_\Delta}$ and Aharonov--Bohm phases $\{h_l\}_{l\in L_\Delta}$ is finite-dimensional and serves as a starting point for the construction of the lattice quantisation. On the lattice, we can introduce a functional Schrödinger representation in terms of complex-valued functionals $\Psi[A]$ of the vector potential. Working on a fixed triangulation $\Delta$, we restrict ourselves to a specific class of such functionals $\Psi[A]$, which are called cylindrical: a functional $\Psi[A]$ of the vector potential is said to be \emph{cylindrical} with respect to the triangulation $\Delta$, symbolically written as $\Psi\in\mathrm{Cyl}_\Delta$, if there exists a square integrable function $f_\Psi:U(1)^N\rightarrow\C$ such that
\begin{equation}
\Psi[A]\equiv( A|\Psi\rangle=f_\Psi\big(h_{l_1}[A],h_{l_2}[A],\dots h_{l_N}[A]\big).
\end{equation}
Between any two such functionals, we have a natural inner product, which is given by
\begin{equation}
\langle \Psi|\Psi'\rangle = \int_{[0,2\pi]^N}\frac{\di\varphi_1\cdots \di\varphi_N}{(2\pi)^N}\overline{f_\Psi(\E^{-\I\varphi_1},\dots,\E^{-\I\varphi_N})}f_{\Psi'}(\E^{-\I\varphi_1},\dots,\E^{-\I\varphi_N}).\label{innprod-def}
\end{equation}
A natural orthonormal basis in the Hilbert space $\mathrm{Cyl}_\Delta$ is given by the charge network states
\begin{equation}
(A|n_1,\dots,n_N\rangle=(h_{l_1}[A])^{n_1}(h_{l_2}[A])^{n_2}\cdots (h_{l_N}[A]\big)^{n_N}.
\end{equation}
The action of the elementary operators is
\begin{align}
\hat{h}_l|n_1,\dots n_N\rangle&=|n_1,\dots n_l+1,\dots n_N\rangle,\\
\hat{\mathcal{E}}^l|n_1,\dots n_N\rangle&=-\hbar e^2 n_l|n_1,\dots n_N\rangle.
\end{align}

The Gauss law \eref{Gsslaw} can now readily be imposed at the quantum level. The constraint is diagonal with respect to the charge network states and it implies that at each node (dual to a tetrahedron $T$) the total electric flux vanishes,
\begin{equation}
\forall T:\sum_{l:t(l)\subset\partial T}n_{l}\,\varepsilon(l,\partial T)=0.
\end{equation}
In here, $\varepsilon(l,\partial T)$ denotes the relative orientation between the link $l$, which lies dual to the triangle $t(l)$, and the boundary of the tetrahedron $T$: $\varepsilon(l,\partial T)=\pm 1$ depending on whether the link enters (leaves) the tetrahedon $T$ and  $\varepsilon(l,\partial T)=0$ in all other cases.\smallskip

The entire construction can be naturally extended to an arbitrarily fine network of links, thereby defining the continuum limit \cite{Ashtekar:1995zh,Ashtekar:1994mh,Fleischhack:2004jc,ALvacuum,Dittrich:2014ala,Asante2023}.
The resulting Hilbert space finds applications in quantum gravity, where it has been conjectured that $U(1)^3$ charge network states can model non-perturbative effects in the $G_{\text{Newton}}\rightarrow 0$ regime of the theory \cite{Thiemann:2022all,Varadarajan:2018tei,Bakhoda:2020ril,Bakhoda:2024mth,Bakhoda:2020fiy,Zarate:2025erv,Smolin:1992wj}. \smallskip

\section{Subsystem phase space for the Maxwell field on local lightfronts}\label{sec3}
\noindent 
This section lays out preparatory material for our main result to be developed in \hyperref[sec4]{Section 4} and \hyperref[sec5]{Section 5}, where we will consider the Poisson algebra of $U(1)$ holonomies on a lightlike (null) initial surface and propose a quantisation of the resulting $U(1)$ holonomy algebra.
\subsection{Newman--Penrose tetrad}
\noindent To compute the Poisson algebra of $U(1)$ holonomies on a null hypersurface $\mathcal{N}$, we choose a future oriented null vector field ${\ell}^{a}$ on $\mathcal{N}$ and extend it into a Newman--Penrose \cite{newmanpenrose,penroserindler} null tetrad $(\vo{4}{k}{a},\vo{4}{\ell}{a},\vo{4}{m}{a},\vo{4}{\bar{m}}{a})$ in a neighbourhood of $\mathcal{N}$. Locally, this is always possible and provides us a tangent space basis, where both $\vo{4}{k}{a}$ and $\vo{4}{\ell}{a}:\vo{4}{\ell}{a}\big|_{\mathcal{N}}=\ell^a\in T\mathcal{N}$ are future oriented null vectors and $\vo{4}{m}{a}$ ($\vo{4}{\bar{m}}{a}$) is spacelike but complex-valued. The only non-vanishing contractions between the elements of this basis are given by
\begin{align}
\vo{4}{k}{a}\vu{4}{\ell}{a}=-1,\qquad \vo{4}{m}{a}\vu{4}{\bar{m}}{a}=1.\label{NP1}
\end{align}
The orientation is fixed by the requirement
\begin{equation}
\varepsilon_{abcd}\vo{4}{\ell}{a}\vo{4}{k}{b}\vo{4}{m}{c}\vo{4}{\bar{m}}{d}=\I.\label{NP2}
\end{equation}
For simplicity and definiteness of the problem, we restrict ourselves to the case in which $\mathcal{N}$ has the topology of a compact interval times a topological two-sphere, $[-1,+1]\times S_2$ with the upper boundary (corner) $\mathcal{C}_+=\{+1\}\times S_2$ lying in the causal future of the lower $\mathcal{C}_-=\{-1\}\times S_2$. By performing a Lorentz transformation $(\vo{4}{k}{a},\vo{4}{\ell}{a},\vo{4}{m}{a},\vo{4}{\bar{m}}{a})\rightarrow (\vo{4}{k}{a}+f\vo{4}{\bar{m}}{a}+\bar{f}\vo{4}{m}{a},\vo{4}{\ell}{a},\vo{4}{m}{a}+f\vo{4}{\ell}{a},\vo{4}{\bar{m}}{a}+\bar{f}\vo{4}{\ell}{a})$, we can then also always achieve
\begin{equation}
\varphi^\ast_{\mathcal{C}_\pm}[\vu{4}{k}{}]_a=0,\label{NP3}
\end{equation}
where $\varphi_{\mathcal{C}_\pm}:\mathcal{C}_\pm\hookrightarrow\mathcal{M}$ is the embedding of the upper (lower) corner into spacetime. If none of the light rays cross or leave the null surface, there is a canonical projection $\pi:\mathcal{N}\rightarrow{\mathcal{C}}_-$, such that the pre-image of every $x\in\mathcal{C}_-$ is the unique light ray that emanates from $x\in\mathcal{C}_-$ and lies tangential to $\mathcal{N}$.  To simplify our language, we call the images of this map \emph{shadows}: if $\mathcal{D}$ is some subset $\mathcal{D}\subset\mathcal{N}$, we call $\pi(\mathcal{D})$ its shadow on $\mathcal{C}_-$, see \hyperref[fig1]{Figure 1} for an illustration. 

\paragraph{Anholonomic coefficients} In the following, we restrict ourselves to such tetrads for which \eref{NP1}, \eref{NP2} and \eref{NP3} are satisfied and, in addition, both $\vo{4}{\ell}{a}$ and $\vo{4}{k}{a}$ are surface orthogonal, such that we are in the same setup as in e.g.\ \cite{Wieland:2020gno}. Thus, the two null vectors define a double null foliation in the neighborhood of $\mathcal{N}$. This implies 
\begin{equation}
\exists \psi_a:\nabla_{[a}\vu{4}{\ell}{b]}=-\psi_{[a}\vu{4}{\ell}{b]},\qquad \varphi^\ast_{\mathcal{N}}[\vu{4}{\ell}{}]_a=0,
\end{equation}
and equally for $\vu{4}{k}{a}$. Without loss of generality, we can now always send $(\vo{4}{k}{a},\vo{4}{\ell}{a})$ into $(\E^\lambda\vo{4}{k}{a},\E^{-\lambda}\vo{4}{\ell}{a})$ such that the relative normalisation \eref{NP1} is maintained but
\begin{equation}
\vu{4}{k}{a}=-\nabla_a u,
\end{equation}
for some clock variable $u$.  The gauge condition \eref{NP3} implies that the upper (lower) corner is a $u=\mathrm{const}.$ surface. By performing a field redefinition $u\rightarrow a u-b$, where $a$ and $b$ are constants, it is then always possible to fix the clock variable such that
\begin{equation}
u\big|_{\mathcal{C}_\pm}=\pm 1.
\end{equation}

For a a given null foliation, it is always possible to find such a co-vector field $\vu{4}{k}{a}$. However, it is only in very special cases that both $\vu{4}{k}{a}$ and $\vu{4}{\ell}{a}$ are both exact one-forms. If we want to maintain ${}^4k=-\di u$, we will have
\begin{equation}
\di\vu{4}{k}{}=0,\quad\text{but}\quad\di\vu{4}{\ell}{}=-\psi\wedge\vu{4}{\ell}{}.\label{kl-ex-deriv}
\end{equation}
The $\vu{4}{\ell}{a}$-component of the one-form $\psi_a$ has a simple geometric interpretation. It determines the inaffinity $\kappa_{(\ell)}$,
\begin{equation}
\vo{4}{\ell}{b}\nabla_b\vo{4}{\ell}{a}=\kappa_{(\ell)}\vo{4}{\ell}{a},\qquad \kappa_{(\ell)}=-\vo{4}{\ell}{a}\psi_a.
\end{equation}
Besides the exterior derivatives \eref{kl-ex-deriv} of $\vu{4}{\ell}{a}$ and $\vu{4}{k}{a}$, we also have the exterior derivatives of $\vu{4}{m}{a}$ (and $\vu{4}{\bar{m}}{a}$). A straight-forward calculation gives
\begin{align}
\di\vu{4}{m}{}&=-\frac{1}{2}\vartheta_{(k)}\vu{4}{\ell}{}\wedge\vu{4}{m}{}-\frac{1}{2}\vartheta_{(\ell)}\vu{4}{k}{}\wedge\vu{4}{m}{}-\I\,\vu{4}{\Gamma}{}\wedge\vu{4}{m}{}+\nonumber\\
&\quad-\sigma_{(\ell)}\vu{4}{k}{}\wedge\vu{4}{\bar{m}}{}-\sigma_{(k)}\vu{4}{\ell}{}\wedge\vu{4}{\bar{m}}{}+\tau\vu{4}{k}{}\wedge\vu{4}{\ell}{}.\label{dm-NP}
\end{align}
In here, $\vartheta_{(\ell)}$ and $\vartheta_{(k)}$ are the expansion of the null congruences formed by $\vo{4}{\ell}{a}$ and $\vo{4}{k}{a}$, while $\sigma_{(\ell)}$ determines the respective shear tensor, i.e\ $\sigma_{(\ell)}=\vo{4}{m}{a}\vo{4}{m}{b}\nabla_a\vu{4}{\ell}{b}$ and equally for $\sigma_{(k)}$. In addition, there is the one-form $\vu{4}{\Gamma}{}$, which transforms as a $U(1)$ spin connection under $U(1)$ frame rotations $\vu{4}{m}{}\rightarrow\E^{\I\lambda}\vu{4}{m}{}$ and $\tau=\vu{4}{m}{a}[\vo{4}{k}{},\vo{4}{\ell}{}]^a$, with $[\cdot,\cdot]$ denoting the Lie bracket of vector fields. 
\paragraph{Hodge dual of two-forms} Given a Newman--Penrose tetrad, we also have an associated basis in the space of two-forms in a neighbourhood of $\mathcal{N}$,
\begin{equation}
\vo{4}{k}{}\wedge\vo{4}{\ell}{},\quad\vo{4}{m}{}\wedge\vo{4}{\bar{m}}{},\quad
\vo{4}{k}{}\wedge\vo{4}{m}{},\quad \vo{4}{k}{}\wedge\vo{4}{\bar{m}}{},\quad
\vo{4}{\ell}{}\wedge\vo{4}{m}{},\quad \vo{4}{\ell}{}\wedge\vo{4}{\bar{m}}{}.
\end{equation}
To conclude this section, we note the Hodge dual ($\star\star=-1$) of these two-forms satisfies
\begin{align}
\star[\vo{4}{k}{}\wedge\vo{4}{\ell}{}]=+\I\vo{4}{m}{}\wedge\vo{4}{\bar{m}}{},\quad\star[\vo{4}{k}{}\wedge\vo{4}{m}{}]=-\I\vo{4}{k}{}\wedge\vo{4}{m}{},\quad\star[\vo{4}{\ell}{}\wedge\vo{4}{m}{}]=+\I\vo{4}{\ell}{}\wedge\vo{4}{m}{},\label{two-frms-dual}
\end{align}
thus, $\vo{4}{\ell}{}\wedge\vo{4}{m}{}$ ($\vo{4}{k}{}\wedge\vo{4}{m}{}$) are (anti)selfdual.
\subsection{Null symplectic structure and Poisson brackets on $\mathcal{N}$}
\noindent 
In this section, we consider the pull-back of the symplectic current \eref{sympl-curr} to a partial Cauchy surface $\mathcal{N}$, which is null. We then compute the resulting symplectic two-form from which we infer the Poisson brackets for the radiative modes of the Maxwell field. 

\paragraph{Vector potential on $\mathcal{N}$} In a neighbourhood of $\mathcal{N}$, we can split the vector potential $\vu{4}{A}{a}$ into the elements of the Newman--Penrose co-basis. Moving forward, we can then also collect the transversal $\di u$-components into an overall $U(1)$ gauge element $\lambda$. This is achieved by integrating the vector potential along the integral curves of $\vo{4}{\ell}{a}$. In this way, we obtain a gauge parameter $\lambda$ that allows us to write
\begin{equation}
\vu{4}{A}{a} = V\vu{4}{\ell}{a}+\bar{\alpha}\vu{4}{m}{a}+\alpha\vu{4}{\bar{m}}{a}+\nabla_a\lambda,\label{A-gauge}
\end{equation}
in a neighbourhood of $\mathcal{N}$. 
We will see below that $(\alpha,\bar{\alpha})$ describe the two radiative modes of the electromagnetic field, whereas $V$ (Coulomb potential) and $\lambda$ (gauge parameter) are edge modes \cite{Freidel:2023bnj,Balachandran:1994up,Carlip:1996yb,Freidel:2015gpa,Freidel:2020xyx,Freidel:2020svx,Freidel:2020ayo,Speranza:2017gxd,Gomes:2016mwl,Wieland:2017zkf,Wieland:2017cmf,Wieland:2021vef,Freidel:2021cjp,Freidel:2021fxf,Carrozza:2021gju,Kabel:2023jve,Giesel:2024xtb,Assanioussi:2023jyq,Langenscheidt:2024nyw,Neri:2025fsh} that enter the symplectic potential through co-dimension two corner terms.\smallskip
  
At the null surface, our basic configuration variable is the pull-back of the vector potential to $\mathcal{N}$,
\begin{equation}
A_a=\varphi^\ast_{\mathcal{N}}\vu{4}{A}{a},\label{N-vec-pot}
\end{equation}
where $\varphi_{\mathcal{N}}:\mathcal{N}\hookrightarrow\mathcal{M}$ is the embedding of the three-dimensional null surface into four-dimensional spacetime. Given the Newman--Penrose tetrad in a neighbourhood of $\mathcal{N}$, we can then
also introduce an associate co-tangent basis in $T^\ast\mathcal{N}$, 
\begin{equation}
k_a=\varphi^\ast_{\mathcal{N}}\vu{4}{k}{a}=-\partial_a u,\quad m_a=\varphi^\ast_{\mathcal{N}}\vu{4}{m}{a},\quad \bar{m}_a=\varphi^\ast_{\mathcal{N}}\vu{4}{\bar{m}}{a},\label{NP-Ntriad}
\end{equation}
On $\mathcal{N}$, we thus have
\begin{equation}
A_a = \bar{\alpha}m_{a}+\alpha\bar{m}_{a}+\partial_a\lambda,\label{N-Apot}
\end{equation}
where $\partial_a$ is the ordinary partial derivative on $\mathcal{N}$. 

\paragraph{Faraday tensor on $\mathcal{N}$} The Faraday tensor is the exterior derivative of the vector potential. Given the gauge conditions \eref{A-gauge}, we obtain
\begin{align}
\vu{4}{F}{}=\di\vu{4}{A}{}
&= \di V\wedge\vu{4}{\ell}{}-V\,\psi\wedge\vu{4}{\ell}{}+\nonumber\\
&\quad-\left(\mathcal{L}_\ell\alpha\,\vu{4}{k}{}\wedge\vu{4}{\bar{m}}{}+\mathcal{L}_k\alpha\,\vu{4}{\ell}{}\wedge\vu{4}{\bar{m}}{}+\CC\right)+\nonumber\\
&\quad-\left(\mathcal{L}_{\bar{m}}\,\alpha\vu{4}{m}{}\wedge\vu{4}{\bar{m}}{}+\alpha\,\di\vu{4}{\bar{m}}{}+\CC\right),\label{4dF-NP}
\end{align}
where e.g.\ $\mathcal{L}_\ell\alpha=\vo{4}{\ell}{a}\nabla_a\alpha$ is the Lie derivative along the four-vector $\vo{4}{\ell}{a}$ of the $\bar{m}$-component (i.e.\ $\alpha$) of the four-vector potential and the symbol \qq{$\CC$} denotes the complex conjugate of all preceding terms in the bracket.\smallskip

Next, we compute the pull-back to the null hypersurface. Going back to the exterior derivative of $\vu{4}{m}{}$ ($\vu{4}{\bar{m}}{}$), i.e.\ \eref{dm-NP}, we obtain, first of all,
\begin{align}
\di m&= -\I\,\Gamma\wedge m-\frac{1}{2}\vartheta_{(\ell)}k\wedge m+\sigma_{(\ell)}k\wedge\bar{m},
\end{align}
where $\Gamma$ transforms as a $U(1)$ spin connection on $\mathcal{N}$. Moving forward, we introduce its components with respect to the co-triad \eref{NP-Ntriad} on $\mathcal{N}$, 
\begin{equation}
\Gamma=\varphi^\ast_{\mathcal{N}}\vu{4}{\Gamma}{}=\varphi_{(\ell)}k+\gamma\bar{m}+\bar{\gamma}m.\label{Gamma-def}
\end{equation}
It is now immediate to compute the pull-back of the Faraday tensor to the null surface. Going back to \eref{4dF-NP} and \eref{N-vec-pot}, we obtain
\begin{align}
F=\di A &=\varphi^\ast_{\mathcal{N}}\vu{4}{F}{}=\nonumber\\
&=+\left(L_{\bar{m}}\alpha+\I\bar{\gamma}\alpha-\CC\right)m\wedge\bar{m}\nonumber\\
&\quad-\left[\left(L_\ell\alpha+\left(\tfrac{1}{2}\vartheta_{(\ell)}-\I\varphi_{(\ell)}\right)\alpha+\sigma_{(\ell)}\bar{\alpha}\right)k\wedge\bar{m}+\CC\right] ,\label{FonN}
\end{align}
where $L_\ell$ and $L_{m}$ ($L_{\bar{m}}$) denotes the Lie derivative on $\mathcal{N}$ with respect to the tangent space basis $(\ell^a,m^a,\bar{m}^a)$ dual to the co-basis $(k_a,m_a,\bar{m}_a)$, i.e.\ $k_a\ell^a=-1$, $m_a\bar{m}^a=1$ with all other contractions vanishing.\smallskip

Next, we consider the pull-back of the Hodge dual of the Faraday tensor. Taking into account the Hodge dual of the elementary two-forms \eref{two-frms-dual}, we obtain
\begin{align}
\star {}F:=\varphi^\ast_{\mathcal{N}}[\star {}^4F]&=-\I{\left(L_\ell V+\kappa_{(\ell)}V-(\bar{\tau}\alpha+\tau\bar{\alpha})\right)}m\wedge\bar{m}+\nonumber\\
&\quad-\I\left[\left(L_\ell\alpha+\left(\tfrac{1}{2}\vartheta_{(\ell)}-\I\,\varphi_{(\ell)}\right)\alpha+\sigma_{(\ell)}\bar{\alpha}\right)k\wedge\bar{m}-\CC\right].\label{starFonN}
\end{align}
Notice the duality between the second line of \eref{starFonN} and \eref{FonN}. Indeed, $\Phi_0=L_\ell\alpha+\big(\tfrac{1}{2}\vartheta_{(\ell)}-\I\varphi_{(\ell)}\big)\alpha+\sigma_{(\ell)}\bar{\alpha}$ is the $k\wedge \bar{m}$ component of the selfdual part $\tfrac{1}{2}(\star +\I)F$ of the Faraday tensor. This observation has important consequences for the Hamiltonian analysis. On a spacelike hypersurface, the electric field and the magnetic vector potential are independent data: up to a $U(1)$ gauge transformation, $E^a$ determines the first time derivative of $A_a$. On a null hypersurface, they are no longer independent initial data, and this leads to the non-commutativity of the vector potential under the Poisson bracket.

\paragraph{Symplectic two-form on $\mathcal{N}$} To equip the boundary data \eref{N-vec-pot} with a  symplectic potential, we integrate the symplectic current \eref{sympl-curr} over the null hypersurface $\mathcal{N}$. Taking into account the field equations $\di\star\vu{4}{F}{}=0$, and $\varphi^\ast_{\mathcal{N}}\vu{4}{\ell}{}=0$, we obtain
\begin{align}
\Theta_{\mathcal{N}}&=-\frac{1}{e^2}\int_{\mathcal{N}}\star F\wedge\left(\bbvar{d}\alpha\,\bar{m}+\bbvar{d}\bar{\alpha}\,{m}+\di(\bbvar{d}\lambda)\right)=\nonumber\\
&=-\frac{1}{e^2}\int_{\mathcal{N}}\Big[\I\left(L_\ell\alpha+\left(\tfrac{1}{2}\vartheta_{(\ell)}-\I\,\varphi_{(\ell)}\right)\alpha+\sigma_{(\ell)}\bar{\alpha}\right)\bbvar{d}\bar{\alpha}\,k\wedge m\wedge\bar{m}+\CC\Big]+\nonumber\\
&\quad+\frac{\I}{e^2}\int_{\partial\mathcal{N}}\Big[L_\ell V+\kappa_{(\ell)}V-(\bar\tau\alpha+\tau\bar{\alpha})\Big]m\wedge\bar{m}\,\bbvar{d}\lambda,\label{sympl-pot-N}
\end{align}
where, going from the first to the second line, we performed a partial integration and used the field equations. Next, we split the integral into its bulk and boundary contributions,
\begin{align}
\Theta_{\mathcal{N}}=\Theta_{\mathcal{N}}^{\mtext{rad}}+\Theta_{\mathcal{C}_+}^{\mtext{edge}}-\Theta_{\mathcal{C}_-}^{\mtext{edge}},
\end{align}
where we introduced the radiative symplectic potential and the edge mode contributions
\begin{align}
\Theta_{\mathcal{N}}^{\mtext{rad}}&=-\frac{1}{e^2}\int_{\mathcal{N}}\Big[\I\left(L_\ell\alpha+\left(\tfrac{1}{2}\vartheta_{(\ell)}-\I\,\varphi_{(\ell)}\right)\alpha+\sigma_{(\ell)}\bar{\alpha}\right)\bbvar{d}\bar{\alpha}\,k\wedge m\wedge\bar{m}+\CC\Big],\label{theta-rad-N}\\
\Theta_{\mathcal{C}}^{\mtext{edge}}&=\frac{\I}{e^2}\oint_{\mathcal{C}}\Big[L_\ell V+\kappa_{(\ell)}V-(\bar\tau\alpha+\tau\bar{\alpha})\Big]m\wedge\bar{m}\,\bbvar{d}\lambda.\label{theta-edge-N}
\end{align}
In here, $E=L_\ell V+\kappa_{(\ell)}V-(\bar\tau\alpha+\tau\bar{\alpha})$ is the electric field on $\mathcal{C}$, which is conjugate to the $U(1)$ edge mode $\lambda$. At the quantum level, the edge mode $\lambda$ serves as a (classical) reference frame for $U(1)$ frame rotations---different values of $\lambda$ correspond to different choices of gauge \cite{Fewster:2025ijg,Donnelly:2016auv,Carrozza:2021gju,Kabel:2023jve}.
\paragraph{Poisson brackets on $\mathcal{N}$} In what follows, we ignore the edge mode contributions to the symplectic structure and consider the Poisson algebra of the radiative modes alone. At the level of the covariant phase space, this can be achieved by introducing a Dirac bracket with respect to some auxiliary constraints that set the values of the electric field $E$ and the gauge parameter $\lambda$ to some prescribed background values $E_o$ and $\lambda_o$, see \cite{Wieland:2021eth}. The constraints $E=E_o$ and $\lambda=\lambda_o$ are second class, the pre-symplectic two-form for the resulting Dirac bracket is then simply given by
\begin{equation}
\Omega_{\mathcal{N}}^{\mtext{rad}}=\bbvar{d}\Theta_{\mathcal{N}}^{\mtext{rad}},
\end{equation}
in which $\bbvar{d}$ denotes the anti-commuting exterior derivative on the space of initial data for the vector potential on $\mathcal{N}$. Since $\bbvar{d}\alpha(p)\,\bbvar{d}\alpha(q)=-\bbvar{d}\alpha(q)\,\bbvar{d}\alpha(p)$, for all $p,q\in\mathcal{N}$, the two terms in \eref{theta-rad-N} that contain the components $\sigma_{(\ell)}$ and $\bar{\sigma}_{(\ell)}$ of the shear tensor $\sigma_{ab}=\sigma_{(\ell)}\bar{m}_a\bar{m}_b+\bar{\sigma}_{(\ell)}m_a{m}_b$ drop out of the symplectic two-form. We obtain
\begin{equation}
\Omega_{\mathcal{N}}^{\mtext{rad}}=-\frac{1}{e^2}\int_{\mathcal{N}}\Big[\I\left(L_\ell(\bbvar{d}\alpha)+\left(\tfrac{1}{2}\vartheta_{(\ell)}-\I\,\varphi_{(\ell)}\right)\bbvar{d}\alpha\right)\bbvar{d}\bar{\alpha}\,k\wedge m\wedge\bar{m}+\CC\Big].\label{Om-N1}
\end{equation}

\paragraph{Coordinates on $\mathcal{N}$} To compute the corresponding Poisson brackets explicitly, and following earlier research on this topic \cite{Wieland:2021vef,Wieland:2024dop,Wieland:2025qgx}, we introduce an explicit coordinate chart and parametrisation of the signature $(0$$+$$+)$ metric on $\mathcal{N}$. As mentioned earlier, let us first note that $k=-\di u$ on $\mathcal{N}$. The clock coordinate $u:\mathcal{N}\rightarrow [-1,1]$ serves as our time variable on $\mathcal{N}$. To complete the coordinate system, we pick a round two-metric\footnote{In the following, indices decorated as $\ubar{a},\ubar{b},\dots$ label tensor fields intrinsic to $\mathcal{C}_-$, i.e.\ sections of $T^r_s\mathcal{C}_-=\bigotimes^rT\mathcal{C}_-\bigotimes^sT^\ast\mathcal{C}_-$.} $\vu{o}{q}{\ubar{a}\ubar{b}}$ on $\mathcal{C}_-$. We can then choose stereographic coordinates $(z,\bar{z})$ on $\mathcal{C}_-$ such that
\begin{equation}
\vu{o}{q}{\ubar{a}\ubar{b}}=2\,\vu{o}{m}{(\ubar{a}}\vu{o}{\bar{m}}{\ubar{b})},\quad\text{for}\quad \vu{o}{m}{\ubar{a}}=\frac{\sqrt{2}}{1+z\bar{z}}\partial_{\ubar{a}}z.\label{fid-mtrc}
\end{equation}
By demanding $\ell^a\partial_az=0$, we can then lift $(z,\bar{z})$ into $\mathcal{N}$. In this way, we obtain a coordinate system $(u,z,\bar{z})$ intrinsic to the null hypersurface $\mathcal{N}$. The tangent vectors $\partial^a_z$, $\partial^a_{\bar{z}}\in T\mathcal{N}$ are spacelike, whereas
\begin{equation}
\ell^a=\partial^a_u\in T\mathcal{N}
\end{equation}
is null. We introduce the fiducial volume elements
\begin{align}
d^3v_o&=-\I\,\di u\wedge\vo{o}{m}{}\wedge\vo{o}{\bar{m}}{},\label{fid-vol3}\\
d^2v_o&=-\I\,\vo{o}{m}{}\wedge\vo{o}{\bar{m}}{}.\label{fid-vol2}
\end{align}
Finally, we parametrize the co-basis elements $(m_a,\bar{m}_a)$ on $\mathcal{N}$ by an overall conformal factor $\Omega$ and an element $S:\mathcal{N}\rightarrow SU(1,1)$,
\begin{equation}
\begin{pmatrix}m_a\\\bar{m}_a\end{pmatrix}=\Omega \underbrace{\begin{pmatrix}S^0&\bar{S}^1\\S^1&\bar{S}^0\end{pmatrix}}_{=S\in SU(1,1)}\begin{pmatrix}\vu{o}{m}{a}\\\vu{o}{\bar{m}}{a}\end{pmatrix},\label{m-SOm}
\end{equation}
see \cite{Wieland:2021vef,Wieland:2024dop,Wieland:2025qgx}. In terms of this parametrisation, shear and expansion can be inferred from
\begin{align}
\Omega^{-1}\frac{\di}{\di u}\Omega &=\frac{1}{2}\vartheta_{(\ell)},\\
\frac{\di}{\di u}SS^{-1}&=\begin{pmatrix}\I\,\varphi_{(\ell)}&\sigma_{(\ell)}\\\bar{\sigma}_{(\ell)}&-\I\,\varphi_{(\ell)}\end{pmatrix}.\label{dotS}
\end{align}
\paragraph{Poisson algebra on $\mathcal{N}$} Given the coordinate system $(u,z,\bar{z})$ introduced above, the symplectic two-form \eref{Om-N1} can be then written as
\begin{equation}
\Omega_{\mathcal{N}}^{\mtext{rad}}=-\frac{1}{e^2}\int_{\mathcal{N}}d^3v_o\left[\frac{\di}{\di u}\left(\Omega\,\E^{-\I\,\Delta}\,\bbvar{d}\alpha\right)\,\Omega\,\E^{\I\,\Delta}\,\bbvar{d}\bar{\alpha}+\CC\right],\label{Om-rad}
\end{equation}
where $\E^{-\I\,\Delta}$ is the $U(1)$ parallel transport  with respect to the spin connection $\Gamma$ on $\mathcal{N}$,
\begin{equation}
\E^{-\I\,\Delta(u,z,\bar{z})}=\E^{-\I\int^u_{-1}\di u'\,\varphi_{(\ell)}(u',z,\bar{z})}.\label{hol-Delta-def}
\end{equation}

Equation \eref{Om-rad} amounts to a canonical transformation in which we reabsorb the conformal factor $\Omega$ and the gravitational $U(1)$ holonomy $\E^{-\I\Delta}$ into a redefinition of the canonical variables $\alpha$ and $\bar{\alpha}$. This simple redefinition is possible, because we work on a fixed background geometry, thus $\bbvar{d}\Omega=0$ and $\bbvar{d}\Delta=0$. 
By inverting the symplectic two-form \eref{Om-rad}, we can then obtain the Poisson brackets, which are given by
\begin{align}
\big\{\alpha(u_1,z_1,\bar{z}_1),\bar{\alpha}(u_2,z_2,\bar{z}_2)\big\}=\frac{e^2}{2}\Theta(u_2-u_1)\Omega_1^{-1}\Omega_2^{-1}\E^{+\I(\Delta_1-\Delta_2)}\delta^{(2)}_{12},\label{Poiss-mods}
\end{align}
where $(\Omega_i,\Delta_i)\equiv(\Omega(u_i,z_i,\bar{z}_i),\Delta(u_i,z_i,\bar{z}_i))$ and $\delta^{(2)}_{12}=\delta^{(2)}(z_1-z_2,\bar{z}_1-\bar{z}_2)$ is the two-dimensional Dirac distribution normalised as $\oint_{\mathcal{C}}d^2v_o\,\delta^{(2)}=1$. Finally, $\Theta(u)$ is the skew-symmetric version of the Heaviside step function,
\begin{equation}
\Theta(u)=\begin{cases}+\tfrac{1}{2},u>0,\\-\tfrac{1}{2},u<0.\end{cases}
\end{equation}
In the \hyperref[appdx]{Appendix} to this paper, we briefly explain how to obtain these Poisson brackets from a Fourier expansion of the radiation field $(\alpha,\bar{\alpha})$ on $\mathcal{N}$.

\section{Holonomy algebra on local lightfronts} 
\label{sec4}\noindent In this section, we turn to our main task in this paper and compute the Possion algebra of the $U(1)$ holonomies on a null initial surface. Given an oriented path $\gamma:[0,1]\rightarrow\mathcal{N}$, we introduce the holonomy as the exponential of the Aharonov--Bohm phase,
\begin{equation}
h_\gamma[A]=\E^{-\I\int_\gamma A}.\label{U1-hlnmy}
\end{equation}
For simplicitly and definiteness of the problem, we restrict ourselves to a special class of paths (\emph{hoops}), which consist of two lightlike segments connected by a spacelike path on $\mathcal{N}$.  Paths are constructed as follows (see \hyperref[fig1]{Figure 1} below). Recall that the  null surface has an upper (lower) boundary $\mathcal{C}_\pm$, $\partial\mathcal{N}=\mathcal{C}_+\cup\mathcal{C}_-^{-1}$. Denote by $\pi:\mathcal{N}\rightarrow\mathcal{C}_-$ the projector from $\mathcal{N}$ onto $\mathcal{C}_-$ for which the pre-image of every $x\in\mathcal{C}_-$ is a null ray in $\mathcal{N}$ emanating from $x$. Then, any of our elementary paths (hoops) $\gamma$ splits into lightlike segments $\gamma_{0,1}:[0,1]\rightarrow\mathcal{N}$, i.e.\ $\pi\circ\gamma=\gamma_{0,1}(0)$, and a spacelike segment $\hat{\gamma}:[0,1]\rightarrow\mathcal{N}$ such that
\begin{equation}
\gamma=\gamma_1^{-1}\circ \hat{\gamma}\circ\gamma_0,
\end{equation}
where the symbol $\circ$ denotes the composition of paths and
\begin{align}
\gamma_0(1)&=\hat{\gamma}(0),\\
\gamma_1(1)&=\hat{\gamma}(1).
\end{align}
At $p_0=\gamma_0(1)$ and $p_1=\gamma_1(1)$ there are turning points, where the path $\gamma$ flips from spacelike to null. In general, our paths (hoops) will not be differentiable at such turning points. Everywhere else they are. \hyperref[fig1]{Figure 1} clarifies the geometry. In what follows, we assume that the connecting path $\hat{\gamma}$ is smooth, i.e.\ $\hat{\gamma}\in\mathcal{C}^\infty([0,1]:\mathcal{N})$. Using the parametrisation of the vector potential given in \eref{N-Apot}, we obtain for any such hoop
\begin{equation}
h_{\gamma}[A]=\mathrm{exp}\Big(-\I\int_{\mathrlap{\hat{\gamma}}}\big(\alpha\bar{m}+\bar{\alpha}m\big)\Big)\E^{-\I\left((\lambda\circ\gamma)(1)-(\lambda\circ\gamma)(0)\right)}.
\end{equation}
\begin{figure}[h]
\centering
\includegraphics[width=0.58\textwidth]{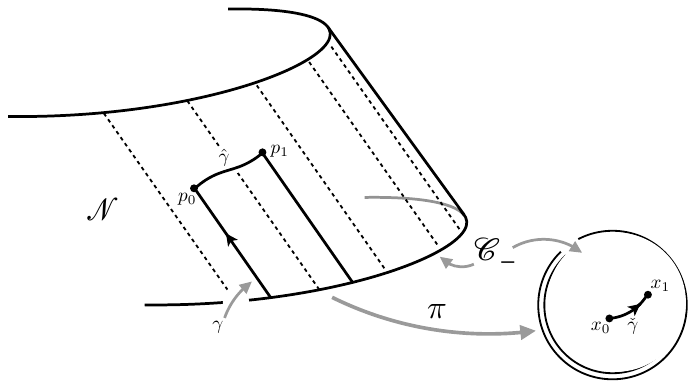}
\caption{We consider $U(1)$ Wilson lines along elementary \emph{hoops} on a null hypersurface $\mathcal{N}$. Any such hoop $\gamma$ consists of two lightlike segments and a spacelike path $\hat{\gamma}$. The corners $p_0$ and $p_1$ are where the different segments join. The projection of the path onto $\mathcal{C}_-$ is $\check{\gamma}=\pi\circ\hat{\gamma}$, where $x_{0,1}=\pi(p_{0,1})$. The images under the projection $\pi$ are what we call \emph{shadows}, i.e.\ $\check{\gamma}=\pi(\hat{\gamma})$ is the shadow of $\hat{\gamma}$. Dotted lines are light rays tangential to $\mathcal{N}$. }\label{fig1}
\end{figure}To compute the Poisson brackets among such Wilson lines, we first note that the gauge parameter $\lambda$ is an edge mode, which Poisson commutes with the radiative modes $\alpha$. In addition, $\lambda$ commutes with itself. For any two such paths $\gamma_+$ and $\gamma_-$, we thus have
\begin{equation}
\big\{h_{\gamma_+},h_{\gamma_-}\big\}=\big\{h_{\hat{\gamma}_+},h_{\hat{\gamma}_-}\big\}\E^{-\I\Delta\lambda},\label{hoop-brackt1}
\end{equation}
where $\hat{\gamma}_\pm$ are the spatial segments of $\gamma_\pm$ and
\begin{equation}
\Delta\lambda=(\lambda\circ\gamma_+)(1)-(\lambda\circ\gamma_+)(0)+(\lambda\circ\gamma_-)(1)-(\lambda\circ\gamma_-)(0).
\end{equation}
\begin{figure}
\centering
\includegraphics[width=0.68\textwidth]{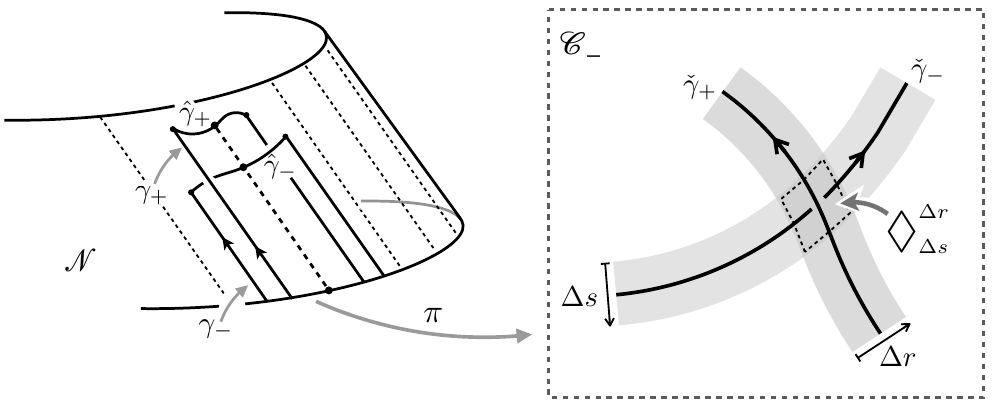}
\caption{The Poisson brackets between two $U(1)$ holonomies for hoops $\gamma_+,\gamma_-\subset\mathcal{N}$ vanish unless the  paths $\gamma_\pm$ intersect with the same light ray (light rays are dashed lines in the left panel). To regularise the Poisson brackets between $U(1)$ holonomies, we extend every path into a two-dimensional ribbon (right panel). The calculation of the Poisson algebra for the smeared $U(1)$ parallel transport  reduces to a two-dimensional integral at the base manifold, i.e.\ the initial cut $\mathcal{C}_-$. The integral vanishes unless the \emph{shadows} of the paths $\hat{\gamma}_\pm$ intersect. }\label{fig2}
\end{figure}We are thus left to compute the Poisson brackets $\{h_{\hat{\gamma}_+},h_{\hat{\gamma}_-}\}$ between the holonomies along the spatial---so to say horizontal---paths $\hat{\gamma}_+, \hat{\gamma}_-\subset\mathcal{N}$, see \hyperref[fig2]{Figure 2} above.
Moving forward, we further simplify our analysis by restricting ourselves to a single crossing between the paths $\check{\gamma}_\pm=\pi\circ\gamma_{\pm}$ at the base manifold, which is the initial corner $\mathcal{C}_-$. We introduce the point $x$ where the paths intersect on the base manifold, and their respective lifts $p_\pm$ to $\hat{\gamma}_{\pm}\cap\mathcal{N}$, i.e.\footnote{In what follows, we use the same symbol for the parameterised paths $\gamma:[0,1]\rightarrow\mathcal{N},t\mapsto\gamma(t)$, and the point sets $\gamma\equiv\{\gamma(t)|t\in[0,1]\}$.}
\begin{equation}
x=\pi(\gamma_{+})\cap\pi(\gamma_{-}), \quad p_\pm=\pi^{-1}(x)\cap\gamma_{\pm}.
\end{equation}
Without loss of generality,\footnote{If the paths $\hat{\gamma}_{\pm}$ intersect in $\mathcal{N}$, the Poisson brackets are ill-defined and depend on giving a meaning to the value $\Theta(0)$ of the step function. A possible assignmenet is $\Theta(0)=0$, in which case $\{h_{\hat{\gamma}_+},h_{\hat{\gamma}_-}\}=0$.} we further assume that $p_+$ lies in the causal future of $p_-$,
\begin{equation}
u_+:=u(p_+)>u(p_-)=:u_-,
\end{equation}
where $u:\mathcal{N}\rightarrow [-1,1]$ is the clock variable introduced above.\smallskip

Next, we smear every path $\hat{\gamma}$ into a thin ribbon $\hat{f}$ thereby introducing a regularisation of the individual holonomy observables,
\begin{equation}
h_{\hat{\gamma}_\pm}=\lim_{\Delta\varepsilon\searrow 0}\mathrm{exp}\Big(-\I\frac{1}{\Delta\varepsilon}\int_{\check{f}_\pm^{\Delta\varepsilon}}\di\varepsilon\wedge \eta^\ast_\pm\left(\alpha\bar{m}+\bar{\alpha}m\right)\Big).
\end{equation}
In here, we anticipated the introduction of several new elements, listed below:
\begin{enumerate}
\item[(i)] \emph{Smearing:} Consider a family of spacelike paths $\{\hat{\gamma}_i\}_{i=1}^N$ on $\mathcal{N}$. In the following, every such path $\hat{\gamma}_i:[0,1]\rightarrow\mathcal{N}$ is smeared into a two-dimensional spacelike surface $\hat{f}^{\Delta\varepsilon}_i\subset\mathcal{N}$, such that the surface $\hat{f}^{\Delta\varepsilon}_i$ (a ribbon) shrinks into the path $\hat{\gamma}_i\subset\mathcal{N}$ as $\Delta\varepsilon\searrow 0$. The projection of the path $\hat{\gamma}_i$ and its smearing surface $\hat{f}^{\Delta\varepsilon}_i$ onto the base manifold (lower corner $\mathcal{C}_-$) are denoted as  $\check{\gamma}_i=\pi(\hat{\gamma}_i)$ and $\check{f}^{\Delta\varepsilon}_i=\pi(\hat{f}^{\Delta\varepsilon}_i)$.
\item[(ii)] \emph{Foliation:} Each surface $\hat{f}^{\Delta\varepsilon}_i\subset\mathcal{N}$ is foliated into a smooth $\varepsilon$-parameter family $\{\hat{\gamma}_{i,\varepsilon}\}_{\varepsilon\in[-\Delta\varepsilon/{2},\Delta\varepsilon/{2}]}$ of paths, $\hat{\gamma}_{i,\varepsilon}:[0,1]\rightarrow\mathcal{N}$. The central trajectory returns the original path $\hat{\gamma}_i$, i.e.\ $\hat{\gamma}_{i,\varepsilon}\big|_{\varepsilon=0}=\hat{\gamma}_i.$
\item[(iii)] \emph{Horizontal lift:} Besides the projection $\pi:\mathcal{N}\rightarrow\mathcal{C}_-$, which is induced by the ruling of the null surface $\mathcal{N}$ by its light rays, we introduce for each surface $\hat{f}^{\Delta\varepsilon}_i$ a corresponding horizontal lift $\eta_i:\mathcal{\mathcal{C}_-}\rightarrow\mathcal{N}$ such that the restriction of $\eta_i\circ\pi$ to $\hat{f}^{\Delta\varepsilon}_i$ is the identity, i.e.\ $\forall p\in\hat{f}^{\Delta\varepsilon}_i:\eta_i(\pi(p))=p$.
\item[(iv)] \emph{Orientation:} The ordered pair of tangent vectors $(\partial^{\ubar{a}}_t,\partial^{\ubar{a}}_\varepsilon)\in T\mathcal{C}_-\times T\mathcal{C}_-$, defined in $\check{f}^{\Delta\varepsilon}_i=\pi(\hat{f}^{\Delta\varepsilon}_i)$ as $\partial^{\ubar{a}}_t\partial_{\ubar{a}} f\big|_{\check{\gamma}_{i,\varepsilon}(t)}=\frac{\di}{\di t}f(\check{\gamma}_{i,\varepsilon}(t))$ and $\partial^{\ubar{a}}_\varepsilon\partial_{\ubar{a}} f\big|_{\check{\gamma}_{i,\varepsilon}(t)}=\frac{\di}{\di \varepsilon}f(\check{\gamma}_{i,\varepsilon}(t))$, for $f:\mathcal{C}_-\rightarrow\R$, is positively oriented in $\mathcal{C}_-$, with $\ubar{a},\ubar{b},\dots$ denoting abstract (co-)tangent space indices on $\mathcal{C}_-$. 
\end{enumerate}
Given these preparations, we can now compute the regularised Poisson bracket, i.e.\
\begin{align}\nonumber
\big\{h_{\hat{\gamma}_+},h_{\hat{\gamma}_-}\big\}&=\lim_{\substack{\Delta r\searrow 0\\\Delta s\searrow 0}}\Big\{\mathrm{exp}\Big(-\I\frac{1}{\Delta r}\int_{\check{f}_+^{\Delta r}}\di r\wedge \eta^\ast_+\left(\alpha\bar{m}+\bar{\alpha}m\right)\Big),\\
&\hspace{6em}\mathrm{exp}\Big(-\I\frac{1}{\Delta s}\int_{\check{f}_-^{\Delta s}}\di s\wedge \eta^\ast_-\left(\alpha\bar{m}+\bar{\alpha}m\right)\Big)\Big\}.\label{Poiss2}
\end{align}
The fundamental Poisson brackets \eref{Poiss-mods} imply that the only non-trivial contributions to this bracket can be reorganized into a two-dimensional integral along the intersection of the smeared paths,
\begin{equation}
\raisebox{-1.2pt}{\scalebox{1.2}[1.2]{$\Diamond$}}^{\Delta r}_{\Delta s}
=\check{f}_+^{\Delta r}\cap \check{f}_-^{\Delta s},
\end{equation}
 see \hyperref[fig2]{Figure 2} for an illustration. Notice now
 \begin{equation}
 \di r\wedge\eta^\ast_\pm \bar{m}=\eta^\ast_\pm (m\wedge \bar{m})\bar{m}^{\ubar{a}}_{\pm}\partial_{\ubar{a}}r=\I\,\eta^\ast_\pm d^2v_o\,\Omega^2_\pm \bar{m}^{\ubar{a}}_\pm\partial_{\ubar{a}}r,
 \end{equation}
where $m^{\ubar{a}}_\pm\in T\mathcal{C}_-$ is the dual to $\eta^\ast_\pm \bar{m}_{\ubar{a}}$, i.e.\ $\eta^\ast_\pm \bar{m}_{\ubar{a}}m^{\ubar{a}}_\pm=1$ and $\eta^\ast_\pm {m}_{\ubar{a}}m^{\ubar{a}}_\pm=0$. In addition, $\Omega_\pm=\Omega\circ\eta_{\pm}$.\smallskip

Taking into account the fundamental Poisson brackets between the radiative modes as given in equation \eref{Poiss-mods}, we obtain
\begin{multline}
\big\{h_{\hat{\gamma}_+},h_{\hat{\gamma}_-}\big\}= \frac{e^2}{4}h_{\hat{\gamma}_+}h_{\hat{\gamma}_-}\lim_{\substack{\Delta r\searrow 0\\\Delta s\searrow 0}}\frac{1}{\Delta r\,\Delta s}\left[+\I\int_{\raisebox{-0.9pt}{\scalebox{0.9}[0.9]{$\Diamond$}}^{\Delta r}_{\Delta s}}\di s\wedge m_-\,\bar{m}^{\ubar{a}}_+\partial_{\ubar{a}}r\right.\\
\times\Omega_+\Omega^{-1}_-\E^{+\I(\Delta_-+\Delta_-)}+\CC\Bigg],\label{Poiss3}
\end{multline}
where $\Delta_\pm=\Delta\circ\eta_\pm$. The integral can be readily evaluated by using the smearing parameters $r$ and $s$, which define the foliation $\gamma_{+,r}$ of $\check{f}^{\Delta r}_+$ and $\gamma_{-,s}$ of $\check{f}^{\Delta s}_-$, as adapted coordinates $(r,s)$ in a neighbourhood of the intersection. Depending on their relative orientation,
\begin{equation}
\mathfrak{o}_{rs}=\operatorname{sign}({}^o\varepsilon_{\ubar{a}\ubar{b}}\partial^{\ubar{a}}_r\partial^{\ubar{b}}_s)=\pm 1,\label{ornt-def}
\end{equation}
we obtain a right-handed (left-handed) coordinate system in a neighbourhood of $x=\pi(\hat{\gamma}_-)\cap\pi(\hat{\gamma}_+)$. In here, ${}^o\varepsilon_{\ubar{a}\ubar{b}}$ denotes the fiducial area element
\begin{equation}
{}^o\varepsilon_{\ubar{a}\ubar{b}}=-2\, \I\,\vu{o}{m}{[\ubar{a}}\vu{o}{\bar{m}}{\ubar{b}]},\qquad \varepsilon_{\ubar{a}\ubar{b}}=\Omega^2\,\vu{o}{\varepsilon}{\ubar{a}\ubar{b}},\label{fid-LCtens}
\end{equation}
see also \eref{fid-mtrc} and \eref{m-SOm}.
In terms of the $(r,s)$-coordinates, the intersection ${\scalebox{1.2}[1.2]{$\Diamond$}}^{\Delta r}_{\Delta s}=\check{f}_+^{\Delta r}\cap \check{f}_-^{\Delta s}$ is simply given by the Cartesian product $[-\Delta r/2,\Delta r/2]\times[-\Delta s/2,\Delta s/2]$. We can now simplify the integral using
\begin{align}\nonumber
\int_{\raisebox{-0.9pt}{\scalebox{0.9}[0.9]{$\Diamond$}}^{\Delta r}_{\Delta s}}\di s\wedge m_-\dots&=\int_{\raisebox{-0.9pt}{\scalebox{0.9}[0.9]{$\Diamond$}}^{\Delta r}_{\Delta s}}\di s\wedge \di r\, [\eta^\ast_-m]_{\ubar{a}}\partial^{\ubar{a}}_r\dots=\\
&=-\int_{-\Delta r/2}^{\Delta r/2}\di r\int_{-\Delta s/2}^{\Delta s/2}\di s\,\mathfrak{o}_{rs}\, [\eta^\ast_-m]_{\ubar{a}}\partial^{\ubar{a}}_r\dots\label{int-simpl1}
\end{align}
Notice now the relationship between the tangent space basis $(\partial^{\ubar{a}}_r,\partial^{\ubar{a}}_s)$ and the dual basis $(\partial_{\ubar{a}}r,\partial_{\ubar{a}}s)$,
\begin{equation}
\partial_{\ubar{a}}r=\frac{\vu{o}{\varepsilon}{\ubar{a}\ubar{b}}\partial^{\ubar{b}}_s}{\vu{o}{\varepsilon}{\ubar{c}\ubar{d}}\partial^{\ubar{c}}_r\partial^{\ubar{d}}_s},\qquad\partial_{\ubar{a}}s=-\frac{\vu{o}{\varepsilon}{\ubar{a}\ubar{b}}\partial^{\ubar{b}}_r}{\vu{o}{\varepsilon}{\ubar{c}\ubar{d}}\partial^{\ubar{c}}_r\partial^{\ubar{d}}_s}.\label{dual-bas}
\end{equation}
Bringing these identities back to \eref{Poiss3}, we obtain
\begin{multline}
\big\{h_{\hat{\gamma}_+},h_{\hat{\gamma}_-}\big\}=-\frac{\I\,e^2}{4}h_{\hat{\gamma}_+}h_{\hat{\gamma}_-}\Bigg[\frac{1}{\big|\vu{o}{\varepsilon}{\ubar{c}\ubar{d}}\partial^{\ubar{c}}_r\partial^{\ubar{d}}_s\big|}[\eta^\ast_-m]_{\ubar{a}}\partial^{\ubar{a}}_r\,[\eta^\ast_+\varepsilon]_{\ubar{b}\ubar{c}}\bar{m}_+^{\ubar{b}}\,\partial^{\ubar{c}}_s\\
\qquad \times\Omega_+^{-1}\Omega_{-}^{-1}\,\E^{+\I(\Delta_+-\Delta_-)}-\CC\Bigg]\Bigg|_x,\label{Poiss4}
\end{multline}
where $x=\pi(\check{\gamma}_+)\cap\pi(\check{\gamma}_-)$ is the intersection of the two paths at the base manifold.
Taking into account that the pull-back of the physical area two-form $\varepsilon_{ab}=-2\I m_{[a}\bar{m}_{b]}$ from $\mathcal{N}$ to $\mathcal{C}_-$ under the horizontal lift $\eta_\pm:\mathcal{C}_-\rightarrow \mathcal{N}$ is
\begin{equation}
[\eta^\ast_+\varepsilon]_{\ubar{a}\ubar{b}}=-2\,\I\, [\eta^\ast_+m]_{[\ubar{a}}[\eta^\ast_+\bar{m}]_{\ubar{b}]},
\end{equation}
and $[\eta^\ast_+m]_{\ubar{a}}m^{\ubar{a}}_+=0,[\eta^\ast_+m]_{\ubar{a}}\bar{m}^{\ubar{a}}_+=1$, we obtain
\begin{align}
\big\{h_{\hat{\gamma}_+},h_{\hat{\gamma}_-}\big\}&=-\frac{e^2}{4}h_{\hat{\gamma}_+}h_{\hat{\gamma}_-}\Bigg[\frac{[\eta^\ast_-m]_{\ubar{a}}[\eta^\ast_+\bar{m}]_{\ubar{b}}\partial^{\ubar{a}}_r\partial^{\ubar{b}}_s}{\big|\vu{o}{\varepsilon}{\ubar{c}\ubar{d}}\partial^{\ubar{c}}_r\partial^{\ubar{d}}_s\big|}\Omega^{-1}_+\Omega^{-1}_-\E^{+\I(\Delta_+-\Delta_-)}+\CC\Bigg]\Bigg|_x.\label{Poiss5}
\end{align}
This expression can be further simplified. First of all, we introduce the two tangent vectors at the intersection $x=\check{\gamma}_+(t_+)=\check{\gamma}_-(t_-)$ of the two paths,
\begin{align}
\check{v}_\pm^{\ubar{a}}=\tfrac{\di}{\di t}\check{\gamma}^{\ubar{a}}_\pm(t_\pm)\in T_x\mathcal{C}_-.
\end{align}
We then also have
\begin{align}
\partial^{\ubar{a}}_r\big|_x=+\mathfrak{o}_{rs}\check{v}^{\ubar{a}}_-,\qquad
\partial^{\ubar{a}}_s\big|_x=-\mathfrak{o}_{rs}\check{v}^{\ubar{a}}_+,
\end{align}
where $\mathfrak{o}_{rs}$ is the orientation \eref{ornt-def}, see \hyperref[fig2]{Figure 2} for an illustration. This implies
\begin{align}
\big\{h_{\hat{\gamma}_+},h_{\hat{\gamma}_-}\big\}&=\frac{e^2}{4}h_{\hat{\gamma}_+}h_{\hat{\gamma}_-}\Bigg[\frac{[\eta^\ast_-m]_{\ubar{a}}[\eta^\ast_+\bar{m}]_{\ubar{b}}\check{v}^{\ubar{a}}_-\check{v}^{\ubar{b}}_+}{\big|\vu{o}{\varepsilon}{\ubar{c}\ubar{d}}\check{v}^{\ubar{c}}_+\check{v}^{\ubar{d}}_-\big|}\Omega^{-1}_+\Omega^{-1}_-\E^{+\I(\Delta_+-\Delta_-)}+\CC\Bigg]\Bigg|_x.\label{Poiss6}
\end{align}

To better understand the geometric meaning of this expression, we introduce the following transport operator
\begin{equation}
\frac{\di}{\di u}\uo{H}{\ubar{a}}{\ubar{b}}=\left(\E^{-2\I\Delta}\sigma_{(\ell)}\partial^{\ubar{b}}_z\partial_{\ubar{c}}\bar{z}+\CC\right)\uo{H}{\ubar{a}}{\ubar{c}}\equiv \uo{{\sigma}}{\ubar{c}}{\ubar{a}}\uo{H}{\ubar{a}}{\ubar{c}},\label{H-def}
\end{equation}
where $\sigma_{(\ell)}$ is the shear, see \eref{dm-NP}, and $\E^{\I\Delta}$ is the $U(1)$ holonomy given in \eref{hol-Delta-def}. The initial condition for integrating this equation is
\begin{equation}
\uo{H}{\ubar{a}}{\ubar{b}}\big|_{u=-1}=\frac{1}{\sqrt{2}}\frac{1+z\bar{z}}{\Omega}\left([\varphi^\ast_{\mathcal{C}_-}m]_{\ubar{a}}\partial^{\ubar{b}}_z+\CC\right)\Big|_{u=-1},
\end{equation}
where $(z,\bar{z})$ are coordinates on $\mathcal{C}_-$, see \eref{fid-mtrc}.\smallskip

Going back to \eref{Poiss6}, we can now write the fundamental Poisson brackets in terms of the holonomy map $\uo{H}{\ubar{a}}{\ubar{b}}$, as introduced in \eref{H-def}, and the fiducial metric $\vu{o}{q}{\ubar{a}\ubar{b}}$, see \eref{fid-mtrc}. Restoring the Poisson brackets for the entire {hoops}, i.e.\ using \eref{hoop-brackt1}, we finally have
\begin{equation}
\big\{h_{\gamma_+},h_{\gamma_-}\big\}=\frac{e^2}{4}h_{\gamma_+}h_{\gamma_-}\frac{\check{v}^{\ubar{a}}_+\check{v}^{\ubar{b}}_-\uo{H}{\ubar{a}}{\ubar{c}}(p_+)\uo{H}{\ubar{b}}{\ubar{d}}(p_-)\vu{o}{q}{\ubar{c}\ubar{d}}}{\big|\vu{o}{\varepsilon}{\ubar{f}\ubar{g}}\check{v}^{\ubar{f}}_+\check{v}^{\ubar{g}}_-\big|}=:\frac{e^2}{4}g(\gamma_+,\gamma_-)h_{\gamma_+}h_{\gamma_-}.\label{Poiss7}
\end{equation}

Let us briefly summarise. \emph{On the left}, we have the Poisson brackets among two paths (hoops) $\gamma_+$ and $\gamma_-$ on the null surface $\mathcal{N}$. As described above, every such path consists of two lightlike segments connected by a spacelike path $\hat{\gamma}_\pm$, see \hyperref[fig1]{Figure 1} and \hyperref[fig2]{2}. The holonomies $h_{\gamma_\pm}$ commute under the Poisson bracket unless the paths $\gamma_+$ and $\gamma_-$ intersect with the same light ray. Here, we assumed one such intersection at $p_\pm=\pi^{-1}(x)\cap\gamma_\pm$, where $x=\pi(\gamma_+)\cap\pi(\gamma_-)$. In addition, we also assumed that $p_+$ lies in the causal future of $p_-$. The generalisation to multiple such intersections and different causal relations is straight-forward. \emph{On the right}, there is the product of holonomies weighted by structure constants $g(\gamma_+,\gamma_-)$. These structure constants depend on the tangent vectors $\check{v}^{\ubar{a}}_\pm\in T\mathcal{C}_-$, which are the push-forward under $\pi_\ast$ of the tangent vectors $\dot{\gamma}^a_\pm\in T\mathcal{N}$ at the intersection between $\gamma_\pm$ and the light ray $\pi^{-1}(x)$. In addition, $\vu{o}{q}{\ubar{a}\ubar{b}}$ and $\vu{o}{\varepsilon}{\ubar{a}\ubar{b}}$ are the fiducial (round) metric \eref{fid-mtrc} and the corresponding Levi-Civita tensor \eref{fid-LCtens}. The tensors ${H}_{\ubar{a}}{}^{\ubar{b}}(p_\pm)$ are the time-ordered exponential of the integral of the shear,
\begin{equation}
\uo{H}{\ubar{a}}{\ubar{b}}(p_\pm)=\tensor{\mathrm{Pexp}\Big(\int_{-1}^{u\pm}\di u\,\boldsymbol{\sigma}(u,z,\bar{z})\Big)}{_{\ubar{a}}^{\ubar{c}}}\uo{H}{\ubar{c}}{\ubar{a}}(x),
\end{equation}
where $\boldsymbol{\sigma}$ stands for the tensor field $\uo{\sigma}{\ubar{a}}{\ubar{b}}\equiv\uo{[\boldsymbol{\sigma}]}{\ubar{a}}{\ubar{b}}$ on $\mathcal{C}_-$, see \eref{H-def}, and $(u_\pm,z,\bar{z})$ are the coordinates of $p_\pm$. Notice also that $\boldsymbol{\sigma}$ is invariant under conformal rescalings of the \emph{physical} boundary metric $q_{ab}\rightarrow \omega^2 q_{ab}$. Therefore, the right hand side of \eref{Poiss7} depends only on the conformal class $[q_{ab}]$ of $q_{ab}=\varphi^\ast_{\mathcal{N}}g_{ab}$. For shear-free null surfaces,\footnote{Immediate examples of such surfaces are isolated horizons \cite{Ashtekar:aa,Ashtekar:2004aa,Ashtekar:2001is,Lewandowski:2002ua,Lewandowski:1999zs} or---even simpler---the e.g.\ $X^0=|\vec{X}|$ or $X^0={X}^3$ surfaces in Minkowski space. }  equation \eref{Poiss7} simplifies further. In that case, $\uo{H}{\ubar{a}}{\ubar{b}}$ is time-independent, and we obtain
\begin{equation}
\text{if}\;\boldsymbol{\sigma}=0:\big\{h_{\gamma_+},h_{\gamma_-}\big\}=\frac{e^2}{4}h_{\gamma_+}h_{\gamma_-}\frac{\cos\varphi({\gamma_+,\gamma_-})}{|\sin\varphi({\gamma_+,\gamma_-})|},\label{Poiss8}
\end{equation}
where $\varphi({\gamma_+,\gamma_-})$ is the angle between the tangent vectors to $\pi(\gamma_+)$ and $\pi(\gamma_-)$ at their intersection. If the two paths intersect in a tangent, i.e.\ $\varphi(\gamma_+,\gamma_-) =0$, the right hand side of \eref{Poiss7} and \eref{Poiss8} diverges. We will see below how the quantum theory can possibly regularise this divergence.

\section{Quantisation}\label{sec5}
\noindent In the previous section, we found the Poisson brackets between $U(1)$ holonomies on lightlike surfaces for classical electromagnetism. This final section lays out a proposal for a representation of the corresponding operator algebra at the quantum level. Quantisation is not an algorithimic process. There are unitarily inequivalent quantum theories that can have the same semi-classical low energy limit.\footnote{The so-called polymer quantisation of a particle on a real line provides a concrete example for such ambiguities. It has the same classical limit as the standard Schrödinger quantisation, but is unitarily inequivalent to it \cite{shadowstats,Ashtekar:2001xp}. The polymer quantisation is based on a represenation of the Weyl algebra, in which the fundamental operators are the exponential of the standard Heisenberg operators. It was developed as a toy model and testing ground for the loop representation of quantum gravity, with which it shares  conceptual similarities \cite{shadowstats,Ashtekar:2001xp,lqcmath}. } In what follows, rather than starting from the standard Poisson algebra of the radiative and the electromagnetic memory modes on $\mathcal{N}$, which could easily be promoted into operators, see \eref{mem-mods} and \eref{rad-mods} in the \hyperref[appdx]{Appendix} below, we consider here directly the Poisson algebra of holonomies alone. The same perspective is taken by the loop representation of quantum geometry, in which the fundamental operators are gravitational holonomies and fluxes smeared over lower-dimensional submanifolds \cite{status,Ashtekar:1994mh,zakolec,Fleischhack:2004jc,ALvacuum,rovelli,ashtekar,thiemann}. \smallskip

We consider a Heisenberg type quantisation. Our starting point is equation \eref{Poiss7}. On the left hand side, we have a Poisson bracket, which can be replaced by $\I/\hbar$ times a commutator. On the right hand side, we obtain an operator product. Using a symmetric ordering, we obtain 
\begin{equation}
\text{for}\;i\neq j:\Big[\hat{h}_{\gamma_i},\hat{h}_{\gamma_j}\Big]=\frac{\hbar e^2}{8\,\I}g_{ij}\left(\hat{h}_{\gamma_i}\hat{h}_{\gamma_j}+\hat{h}_{\gamma_j}\hat{h}_{\gamma_i}\right)\label{QAlgbr1},
\end{equation}
where $\{\gamma_i\}_{i\in I}$ denotes a family of oriented paths on the null surface $\mathcal{N}$, 
where the structure constants $g_{ij}=g(\gamma_i,\gamma_j)$ are defined as in \eref{Poiss7} above. We assume further that none of the paths intersect $\gamma_i\cap\gamma_j=\emptyset$ and that the projection of any pair of paths onto the base manifold has at most one intersection, see \hyperref[fig2]{Figure 2} above. The generalisation to multiple such intersections is straight forward. In what follows, we call the family of paths $\{\gamma_i\}_{i\in I}$ that underlies the construction a \emph{graph}, symbolically denoted as $\Gamma:=\{\gamma_i\}_{i\in I}$\smallskip

In the simple case of a shear-free null surface, such as e.g.\ an isolated horizon, the structure constants are  given by
 \begin{equation}
g_{ij}=\varepsilon_{ij}\frac{\cos\varphi({\gamma_i,\gamma_j})}{|\sin\varphi({\gamma_i,\gamma_j})|}.
\end{equation}
In here, $\varphi({\gamma_i,\gamma_j})$ denotes the angle at which the paths $\{\pi\circ\gamma_i\}_{i\in I}$ intersect at the base manifold and $\varepsilon_{ij}=\pm 1$ depending on whether the intersection $p_i:=\gamma_i\cap\pi^{-1}(\pi(\gamma_i)\cap\pi(\gamma_i))$ lies in the causal future (past) of $p_j$. If there is no intersection $g_{ij}=0$. The projection $\pi:\mathcal{N}\rightarrow\mathcal{C}_-$ is performed with respect to the null rays tangential to $\mathcal{N}$. Every null ray is projected into its intersection with the initial cut $\mathcal{C}_-$, see \hyperref[fig1]{Figure 1} above.\smallskip 

Notice then that equation \eref{QAlgbr1} is the same as to say
\begin{equation}
\text{for}\;i\neq j:\hat{h}_{\gamma_i}\hat{h}_{\gamma_j}=\E^{-2\pi\I q_{ij}}\hat{h}_{\gamma_j}\hat{h}_{\gamma_i}\label{QAlgbr2},
\end{equation}
where we defined the deformation parameter
\begin{equation}
\E^{2\pi\I q_{ij}}\equiv\E^{2\pi\I q(\gamma_i,\gamma_j)}:=\frac{1+\frac{\I\hbar e^2}{8}g_{ij}}{1-\frac{\I\hbar e^2}{8}g_{ij}}.\label{q-angle}
\end{equation}

Image now a smooth family of paths $\{\gamma_\varepsilon\}_{\varepsilon\in[-1,1]}$ such that for all $\varepsilon\neq\varepsilon'$, $\gamma_\varepsilon\cap\gamma_{\varepsilon'}=\emptyset$, but $\pi(\gamma_\varepsilon)\cap\pi(\gamma_{\varepsilon'})=x\in\mathcal{C}_-$. This is the same as to say that none of the paths $\{\gamma_\varepsilon\}_{\varepsilon\in[-1,1]}$ intersect, but they all intersect with the same null ray $\pi^{-1}(x)$. Consider now the coincidence limit $\varepsilon'\rightarrow \varepsilon$ of the classical and quantum structure constants $g(\gamma_\varepsilon,\gamma_{\varepsilon'})$ and $q(\gamma_\varepsilon,\gamma_{\varepsilon'})$. Whereas the classical structure constants $g(\gamma_\varepsilon,\gamma_{\varepsilon'})$ diverge, the quantum mechanical structure constants $q(\gamma_\varepsilon,\gamma_{\varepsilon'})$  are finite,
\begin{equation}
\lim_{\varepsilon'\rightarrow\varepsilon}\E^{2\pi\I q(\gamma_{\varepsilon'},\gamma_\varepsilon)}=-1.
\end{equation}
This suggest to set
\begin{equation}
\hat{h}_\gamma\hat{h}_\gamma=-\hat{h}_\gamma\hat{h}_\gamma=0,
\end{equation}
which implies that each holonomy is replaced by an anti-commuting Grassman variable.\smallskip

To push this idea forward and complete the quantisation, we also need to speak about the reality conditions that determine the inner product. At the classical level, the holonomies take values in $U(1)$ and satisfy the reality conditions
\begin{equation}
h_{\gamma}\bar{h}_{\gamma}=|h_\gamma|^2=1,\qquad \bar{h}_\gamma = h_{\gamma^{-1}},\label{class-real-cond}
\end{equation}
where $\gamma^{-1}$ denotes the inversely oriented path. To represent these algebraic relations at the quantum level, we choose a symmetric ordering
\begin{equation}\label{QAlgbr2}
\frac{1}{2}\left(\hat{h}_{\gamma}\hat{h}^\dagger_{\gamma}+\hat{h}^\dagger_{\gamma}\hat{h}_{\gamma}\right)=\bbvar{1},\qquad \hat{h}^\dagger_\gamma=\hat{h}_{\gamma^{-1}},
\end{equation}
where $\bbvar{1}$ denotes the identity element $\bbvar{1}\hat{h}_{\gamma}=\hat{h}_{\gamma}\bbvar{1}=\hat{h}_{\gamma}$ of the holonomy algebra.
Combining \eref{QAlgbr1} and \eref{QAlgbr2}, we obtain, in this way, the fundamental algebraic relations
\begin{align}
\hat{h}_{\gamma_i}\hat{h}_{\gamma_j}&=\E^{-2\pi\I q_{ij}}\hat{h}_{\gamma_j}\hat{h}_{\gamma_i},\label{QAlgbr3}\\
\hat{h}_{\gamma_i}\hat{h}_{\gamma_j}^\dagger&=\E^{+2\pi\I q_{ij}}\hat{h}_{\gamma_j}^\dagger\hat{h}_{\gamma_i}+2\delta_{ij}\bbvar{1}.\label{QAlgbr4}
\end{align}
To introduce a Hilbert space representation, consider first the algebra generated by a single path $\gamma$. The resulting Hilbert space is two-dimensional and spanned by orthonormal basis vectors $|\gamma,\uparrow\rangle$ and $|\gamma,\downarrow\rangle$ that satisfy
\begin{equation}\left.
\begin{array}{rclrcl}
\hat{h}^\dagger_\gamma|\gamma,\uparrow\rangle&=&0,& \hat{h}_\gamma|\gamma,\uparrow\rangle&=&\sqrt{2}|\gamma,\downarrow\rangle,\\
\hat{h}^\dagger_\gamma|\gamma,\downarrow\rangle&=&\sqrt{2}|\gamma,\uparrow\rangle,& \hat{h}_\gamma|\gamma,\downarrow\rangle&=&0.
\end{array}\right\}
\end{equation}
The reality conditions $h_{\gamma^{-1}}=h_\gamma^\dagger$ imply
\begin{equation}
|\gamma^{-1},\uparrow\rangle=|\gamma,\downarrow\rangle.
\end{equation}
To extend this construction to the entire holonomy algebra, we assign to every oriented graph $\gamma\in\Gamma$ an element of $\{\uparrow,\downarrow\}$ and call any such assignment $\gamma\mapsto s (\gamma)\in\{\uparrow,\downarrow\}$ a framing $s(\gamma)$.   We can then define a vacuum state by declaring that for all subsequences $(\gamma_{\pi(1)},\dots,\gamma_{\pi(K)})$ of $(\gamma_1,\dots,\gamma_N)$\begin{equation}
\hat{h}_{\gamma_{\pi(1)}}^{s(\gamma_{\pi(1)})}\cdots\hat{h}_{\gamma_{\pi(K)}}^{s(\gamma_{\pi(K)})}|s,\Gamma\rangle=0,\quad\text{where}\quad\begin{cases}
\hat{h}_{\gamma}^{\uparrow}=\hat{h}_{\gamma}^\dagger,\\
\hat{h}_{\gamma}^{\downarrow}=\hat{h}_{\gamma},
\end{cases}\label{s-vac}
\end{equation}
and $\Gamma=\{\gamma_i\}_{i=1}^N$ denotes the graph (family of paths) under consideration. Given this state, we can can then use the algebraic relations \eref{QAlgbr3} and \eref{QAlgbr4} to compute the expectation values
\begin{equation}
\langle \hat{x}\rangle=\langle s,\Gamma|\hat{x}|s,\Gamma\rangle.\label{GNS-vac}
\end{equation}
for a generic element $\hat{x}=a\bbvar{1}+\sum_{i} a_i \hat{h}_{\gamma_i}+\sum_i b_i \hat{h}^\dagger_{\gamma_i}+\sum_{ij}a_{ij}\hat{h}_{\gamma_i}\hat{h}_{\gamma_j}+\dots$ of the holonomy algebra, where $a,a_{i},b_{i},a_{ij},\dots\in\C$.
The map $ \hat{x}\rightarrow \langle \hat{x}\rangle$ defines a state on the algebra, which, through the Gelfand–Naimark–Segal (GNS) construction, provides us with a Hilbert space representation of the holonomy algebra \eref{QAlgbr3} and \eref{QAlgbr4} for a fixed family of paths. The entire construction depends on a choice of vacuum state $|s,\Gamma\rangle$, or, which is to say the same thing, a choice of framing $s(\gamma)\in\{\uparrow,\downarrow\}$ for each path $\gamma\in \Gamma$. To make the construction independent of the orientation of $\gamma$, we set $s(\gamma^{-1})=s^{-1}(\gamma)$, where $\uparrow^{-1}=\downarrow$ and $\downarrow^{-1}=\uparrow$.

\paragraph{Cylindrical consistency} So far, we have considered a quantum representation of the holonomy algebra for a \emph{fixed} family of framed paths, i.e.\ a graph $\Gamma=\{\gamma_i\}_{i=1}^N$ consisting of a family of oriented paths $\gamma_i$ equipped with a framing $s(\gamma_i)\in\{\uparrow,\downarrow\}$. Given the state \eref{GNS-vac}, we can construct a corresponding graph Hilbert space $\mathcal{H}_\Gamma$, $\dim(\mathcal{H}_\Gamma)=2^N$, which is invariant under orientation reversal sending $\gamma\in\Gamma$ into $\gamma^{-1}$ and $s(\gamma)$ into $s^{-1}(\gamma)$. Given two graphs, $\Gamma, \Gamma'$, we obtain, in this way, different graph Hilbert spaces $\mathcal{H}_\Gamma$ and $\mathcal{H}_{\Gamma'}$. How are they related?  

Graphs are equipped with a naturally partial order $(\{\Gamma,\Gamma',\dots\},\prec)$.  
If we then replace $\Gamma$ by a finer graph $\Gamma':\Gamma'\prec\Gamma$, there should be an isometry $\iota_{\Gamma\rightarrow\Gamma'}:\mathcal{H}_\Gamma\rightarrow\mathcal{H}_{\Gamma'}$ that embeds the smaller into the larger graph Hilbert space. The construction is said to be \emph{cylindrical consistent} if it is stable under graph refinements, i.e.\ $\forall\Gamma':\iota_{\Gamma'\rightarrow\Gamma''}\circ\iota_{\Gamma\rightarrow \Gamma'}=\iota_{\Gamma\rightarrow\Gamma'}$. The concept is key for constructing diffeomorphism-invariant measures on the space of generalised connections \cite{Ashtekar:1995zh,Ashtekar:1993wf,ALvacuum} and plays an important role in the definition of the continuum limit of the spinfoam approaches to covariant quantum gravity \cite{Asante2023,Dittrich:2014ala}. How can we construct such isomorphisms? 

If $\Gamma$ and $\Gamma'$ are disjoint graphs and $\Gamma''=\Gamma\cup\Gamma'$ denotes the union of graphs, a natural choice for such an isomorphism is merely the tensor product $\iota_{\Gamma\rightarrow\Gamma''}:\mathcal{H}_\Gamma\rightarrow\mathcal{H}_{\Gamma''}=\mathcal{H}_\Gamma\otimes \mathcal{H}_{\Gamma'},|\psi\rangle\mapsto|\psi\rangle\otimes|s,\Gamma'\rangle$ with its natural identification by the vacuum state given in \eref{GNS-vac}. Orientation reversal 
is equally unproblematic. The more difficult part is to explain what happens when we split an individual path $\gamma\in\Gamma$ into component paths $\alpha,\beta:\gamma=\beta\circ\alpha$, thereby replacing $\Gamma=\{\gamma,\gamma',\dots\}$ by a finer graph $\Gamma'=\{\alpha,\beta,\gamma',\dots\}$. At the classical level, path refinements imply gluing conditions $h_\gamma=h_\beta h_\alpha$ between the path holonomies $h_\alpha,h_\beta$ and $h_\gamma$. This follows from the elementary definition of the Aharonov--Bohm phases \eref{U1-hlnmy}. At the quantum level, we can impose the gluing condition as a constraint $\hat{C}=\hat{h}_\gamma-\hat{h}_\beta \hat{h}_\alpha$ on the extended Hilbert space $\mathcal{K}_{\Gamma,\Gamma'}=\mathcal{H}_\Gamma\otimes\mathcal{H}_{\Gamma'}$.\footnote{Notice that the constraints $\hat{C}$ and $\hat{C}^\dagger$ form a second-class system. Depending on the choice of framing of the underlying path $\gamma$, see equation \eref{s-vac} above, physical states $\psi_{\mtext{phys}}\in\mathcal{H}_{\mtext{phys}}$ lie in the kernel of $\hat{C}$ (if $s=\,\downarrow$) or $\hat{C}^\dagger$ (if $s=\,\uparrow$). Here, we restrict ourselves to the case $s=\,\downarrow$. The conjugate operator  $\hat{C}^\dagger$
maps $\mathcal{H}_{\mtext{phys}}$ into its orthogonal complement, $(\hat{C}^{s})^\dagger\mathcal{H}_{\mtext{phys}}\perp\mathcal{H}_{\mtext{phys}}$.} For a single link $\gamma=\beta\circ\alpha$, the states in the kernel of the constraint are then given by
\begin{subequations}
\begin{align}
E_{\downarrow\downarrow}&=|\gamma,\downarrow\rangle\otimes|\beta,\downarrow\rangle\otimes|\alpha,\downarrow\rangle,\\
E_{\uparrow\uparrow}&=\sqrt{\frac{2}{3}}\left(|\gamma,\uparrow\rangle\otimes|\beta,\downarrow\rangle\otimes|\alpha,\downarrow\rangle+\frac{1}{\sqrt{2}}|\gamma,\downarrow\rangle\otimes|\beta,\uparrow\rangle\otimes|\alpha,\uparrow\rangle\right),\\
E_{\downarrow\uparrow}&=|\gamma,\downarrow\rangle\otimes|\beta,\downarrow\rangle\otimes|\alpha,\uparrow\rangle,\\
E_{\uparrow\downarrow}&=|\gamma,\downarrow\rangle\otimes|\beta,\uparrow\rangle\otimes|\alpha,\downarrow\rangle.
\end{align}
\end{subequations}
These basis vectors span a four-dimensional subspace $\mathcal{H}_{\{\gamma\},\{\alpha,\beta\}}$  of $\mathcal{K}_{\{\gamma\},\{\alpha,\beta\}}=\mathcal{H}_{\{\gamma\}}\otimes\mathcal{H}_{\{\alpha,\beta\}}$. The natural embeddings maps of $\mathcal{H}_{\{\gamma\}}$ and $\mathcal{H}_{\{\alpha,\beta\}}$ into $\mathcal{H}_{\{\gamma\},\{\alpha,\beta\}}$ are defined by $\iota_{\{\gamma\}}|\gamma,r\rangle = E_{rr}$ and $\iota_{\{\alpha,\beta\}}\big(|\beta,r\rangle\otimes|\alpha,s\rangle\big) = E_{rs}$ for $r,s\in\{\uparrow,\downarrow\}$. Stability under graph refinements follows then simply from the associativity of the tensor product.
\section{Concluding remarks}
\noindent This article was prepared for the special collection of \emph{General Relativity and Gravitation} on ``Geometry of Classical and Quantum Space-Times'' in memory of Prof.\ Jerzy Le\-wan\-dowski (1959--2024). 
Much of what we have laid out here in this article builds on Prof.\ Le\-wan\-dowski's earlier research and collaborations. 
\hyperref[sec2]{Section 2} is a short review of the charge network representation of the quantum Maxwell field. In here, we restricted ourselves to holonomies on a graph dual to a fixed triangulation. The generalisation to arbitrary graphs has been established by Le\-wan\-dowski and collaborators in \cite{Ashtekar:1994mh,Ashtekar:1995zh,ALvacuum}.  
Our main results are developed in \hyperref[sec3]{Section 3}, \hyperref[sec4]{Section 4} and \hyperref[sec5]{Section 5}. In \hyperref[sec3]{Section 3}, we introduced the phase space of a local subsystem of the Maxwell field on a null initial surface. Further details are given in the \hyperref[appdx]{Appendix} below. 
Many of the technical details of the construction resonate with recent results on non-expanding and isolated horizons \cite{Ashtekar:aa,Ashtekar:2004aa,Ashtekar:2001is,Lewandowski:2002ua,Lewandowski:1999zs,Ashtekar:2021wld,Ashtekar:2024stm,Ashtekar:2021kqj,Ashtekar:2021wld}, but there are also differences. In our case, the null surface may have expansion and shear, yet all the shear terms drop out of the final expression for the symplectic two-form \eref{Om-rad}. The expansion, on the other hand, can be reabsorbed into a redefinition of the canonical variables, see \eref{Om-rad}. The simplification of the symplectic structure was possible, because we kept the gravitational background fixed (the variation of all metric components vanish). The resulting Poisson brackets between the $U(1)$ holonomies were considered in \hyperref[sec4]{Section 4}. Here, our results drastically differ from the standard $U(1)$ holonomy-flux algebra on spacelike initial surfaces. On a spacelike hypersurface, electric and magnetic fluxes each form a commuting subalgebra on phase space. Hence holonomies commute. On a null surface, this is no longer true. This is a direct consequence of $\star(k\wedge m)=-\I\,k\wedge m\in\Omega^2(\mathcal{N}:\C)$, i.e.\ the fact that there are self-dual two-forms that are intrinsic to a null hypersurface such that the pull-back of ${}^4F$ (magnetic fields) and $\star{}^4F$ (electric fields) are no longer independent variables on phase space. This in turn implies that the holonomies (exponentials of magnetic fluxes) become non-commutative. Given two Wilson lines that intersect the same light ray, their Poisson brackets no longer commute. To compute the Poisson brackets explicitly, we introduced a regularisation, in which the underlying paths are smeared into two-dimensional thin ribbons. The structure constants of the resulting holonomy algebra depend on only the conformal class of the signature $(0$$+$$+$$)$ metric at the boundary. The conformal factor drops out of the Poisson brackets. Finally, we developed a proposal (see \hyperref[sec4]{Section 5}) for a quantistaion of the holonomy algebra. This quantisation exhibits a few suprising features. First, the structure constants are regular, whereas the classical Poisson brackets between the Wilson lines are singular when their projection (i.e.\ their shadows) onto a two-dimensional cut of the null surface intersect tangentially, see \hyperref[fig1]{Figure 1} for an illustration. Second, there is no longer a unique ground state. The vacuum state depends on a choice of framing. This is markedly different to a spacelike hypersurface, where there is a unique diffeomorphism invariant ground state of the holonomy flux algebra \cite{ALvacuum}. Third, each individual holonomy behaves as an anti-commuting creation (annihilation) operator. If we take two distinct Wilson lines that intersect the same light ray, we obtain a different behaviour. Depending on the geometry at the intersection, the commutation relations between different path holonomies continuously interpolate between fermionic and bosonic statistics. For the non-Abelian case, not studied in this paper, we expect a much richer algebraic structure possibly related to quantum groups \cite{Majid_1995}. From the perspective of conformal field theories (CFTs) in two dimensions, our results are perhaps not particularly surprising. On each light ray, the symplectic potential \eref{Om-rad} can be mapped into the symplectic potential of an auxiliary conformal field theory (CFT) for a free charged bosonic scalar field $\varphi=\Omega\E^{-\I\Delta}\alpha$.  A similar correspondence exists also for gravity, in which case we have both bosonic and fermionic representations of gravitational current algebras on the light cone \cite{Ciambelli:2023mir,Wieland:2025qgx}. If we then take the exponential of such a bosonic scalar field---which is what we do when computing the holonomy---we obtain a new composite observable, which behaves as a free fermion \cite{CFTbook}.\smallskip


\paragraph{Acknowledgments} W.W.\ thanks Yuki Yokukura for hospitality at KEK (Tsukuba) and Philipp Höhn for hospitality at OIST (Okinawa). This research was funded in parts through the Heisenberg programme of \emph{Deutsche Forschungsgemeinschaft} (DFG, German
Research Foundation)--–543301681.

\appendix
\section*{Appendix: Poisson brackets for the Maxwell field on a null surface interval}\label{appdx}
\renewcommand\theequation{A.\arabic{equation}}
\noindent In this section, we briefly explain how to arrive at the fundamental Poisson brackets \eref{Poiss-mods} for the radiation field. For simplicity, we ignore the angular modes and consider the symplectic potential
\begin{equation}
\Omega=-\frac{1}{e^2}\int_{-1}^1\di u\frac{\di}{\di u}\bbvar{d}\alpha(u)\,\bbvar{d}\bar{\alpha}(u)+\CC
\end{equation}
Because of radiative memory, generic configurations of the radiation field will not satisfy specific fixed boundary conditions such as e.g.\ periodic boundary conditions $\alpha(-1)=\alpha(+1)$. It is therefore useful to split $\alpha(u)$ into two terms: a periodic part $\alpha_{\mtext{hard}}(u)$ and a linear term $\alpha_{\mtext{soft}}(u)$,\footnote{That we refer to these two parts as \emph{hard} and \emph{soft} parts of the radiation field is borrowed from standard terminology at null infinity, see e.g.\ \cite{AshtekarNullInfinity,Strominger:2017zoo,Pasterski:2015zua}.}  
\begin{align}
\alpha(u)=\alpha_{\mtext{soft}}(u)+\alpha_{\mtext{hard}}(u),
\end{align}
where
\begin{subequations}
\begin{align}
\alpha_{\mtext{soft}}(u)&=\frac{1}{2}\Delta\alpha\,u,\\
\alpha_{\mtext{hard}}(u)&=\sum_{n=-\infty}^\infty\alpha_n\E^{-\I n\pi u}.
\end{align}
\end{subequations}
A short calculation gives
\begin{align}
\Omega&=\frac{1}{e^2}\bbvar{d}(\Delta\alpha)\,\bbvar{d}\left(\bar{\alpha}_{\mtext{hard}}(u=1)\right)-\frac{1}{e^2}\int_{-1}^1\di u\,\bbvar{d}(\Delta\alpha)\,\bbvar{d}\bar{\alpha}_{\mtext{hard}}(u)+\nonumber\\
&\quad-\frac{1}{e^2}\int_{-1}^1\di u\,\frac{\di}{\di u}\Big(\bbvar{d}\alpha_{\mtext{hard}}(u)\Big)\bbvar{d}\bar{\alpha}_{\mtext{hard}}(u)+\CC=\nonumber\\
&=-\frac{1}{e^2}\bbvar{d}\left(\Delta\alpha\right)\Big(\bbvar{d}\alpha_0+\sum_{n\neq 0}(-1)^{n+1}\bbvar{d}\bar{\alpha}_n\Big)+\frac{2\I\pi}{e^2}\sum_{n\neq 0}n\,\bbvar{d}\alpha_n\,\bbvar{d}\bar{\alpha}_n+\CC,\label{Om-def-appdx}
\end{align}
where $\CC$ denotes the complex conjugate of \emph{all} preceding terms.  The fundamental Poisson brackets are then given by
\begin{align}
\big\{p_{\Delta\alpha},\Delta\alpha\big\}&=\big\{p_{\Delta\bar{\alpha}},\Delta\bar{\alpha}\big\}=1,\label{mem-mods}\\
\big\{\alpha_n,\bar{\alpha}_m\big\}&=-\frac{\I\,e^2}{4\pi}\frac{1}{n}\delta_{n-m},\qquad n,m\neq 0,\label{rad-mods}
\end{align}
where $p_{\Delta\alpha}$ denotes the momentum conjugate to the finite memory $\Delta\alpha=\alpha(1)-\alpha(-1)$,
\begin{equation}
p_{\Delta\alpha}=\frac{1}{e^2}\Big(\bar{\alpha}_0+\sum_{n\neq 0}(-1)^{n+1}\bar{\alpha}_n\Big).
\end{equation}
This phase space can be readily quantised. The resulting Hilbert space is the tensor product of the Fock space for the oscillators $\{\alpha_n\}_{n\neq 0}$, in which the positive frequency modes act as annihilation operators, and the Hilbert space $\mathcal{H}_{\mtext{soft}}=L^2(\C,\tfrac{1}{2\I}\di(\Delta\alpha)\di(\Delta\bar{\alpha}))$ for the soft modes (electromagnetic memory).\smallskip

Given the fundamental Poisson brackets \eref{mem-mods} and \eref{rad-mods}, it is now straight forward to show
\begin{align}
\big\{\alpha(u_+),\bar{\alpha}(u_-)\big\}&=g(u_+,u_-),
\end{align}
where the dirstribution $g(u_+,u_-)$ is formally given by
\begin{align}
g(u_+,u_-)&=-\frac{e^2}{2}(u_+-u_-)\nonumber\\
&\qquad-\frac{\I\,e^2}{4\pi}\sum_{n\neq 0}\frac{1}{n}\Big(1+\E^{-\I n\pi(u_++1)}+\E^{\I n\pi(u_-+1)}+\E^{-\I n\pi(u_+-u_-)}\Big).
\end{align}
Notice that $g(u_+,u_-)=-g(u_-,u_+)$ and
\begin{equation}
\frac{\di}{\di u_+}g(u_+,u_-)=-\frac{e^2}{2}\delta\big((u_++1)\operatorname{mod} 2\big)-\frac{e^2}{2}\delta\big((u_+-u_-)\operatorname{mod} 2\big),
\end{equation}
where $\delta\big(u\operatorname{mod} 2\big)$ is the Dirac comb with singualrities at $u=0,\pm2,\dots$ We integrate this expression in the interval $-1<u_\pm<1$ obtaining
\begin{equation}
g(u_+,u_-)=c(u_-)-\frac{e^2}{2}\Theta(u_+-u_-),
\end{equation}
where $c(u_-)$ is an integration constant. The requirement $g(u_+,u_-)=-g(u_-,u_+)$ sets $c(u_-)=0$.

\providecommand{\href}[2]{#2}\begingroup\raggedright\endgroup


\begin{thebibliography}{100}

\bibitem{status}
A.~Ashtekar and J.~Lewandowski, ``{Background independent quantum gravity: a
  status report},'' {\em Class. Quant. Grav.} {\bf 21} (2004), no.~15,
  R53--R152, \href{http://arXiv.org/abs/gr-qc/0404018v2}{{\tt
  arXiv:gr-qc/0404018v2}}.

\bibitem{ashtekar}
A.~Ashtekar, {\em {Lectures on Non-Pertubative Canonical Gravity}}.
\newblock World Scientific, 1991.

\bibitem{rovelli}
C.~Rovelli, {\em Quantum Gravity}.
\newblock Cambridge University Press, Cambridge, 2008.

\bibitem{thiemann}
C.~Thiemann, {\em Introduction to Modern Canonical Quantum General Relativity}.
\newblock Cambridge University Press, 2007.

\bibitem{zakolec}
C.~Rovelli, ``{Zakopane lectures on loop gravity},'' {\em PoS} {\bf QGQGS2011}
  (2011) 003,
\href{http://arXiv.org/abs/1102.3660}{{\tt arXiv:1102.3660}}.

\bibitem{Gambini_Pullin_1996}
R.~Gambini and J.~Pullin, {\em Loops, Knots, Gauge Theories and Quantum
  Gravity}.
\newblock Cambridge Monographs on Mathematical Physics. Cambridge University
  Press, 1996.

\bibitem{alexreview}
A.~Perez, ``{The Spin-Foam Approach to Quantum Gravity},'' {\em Living Rev.
  Rel.} {\bf 16} (2013), no.~3,
\href{http://arXiv.org/abs/1205.2019}{{\tt arXiv:1205.2019}}.

\bibitem{Rovelli:2014ssa}
C.~Rovelli and F.~Vidotto, {\em {Covariant Loop Quantum Gravity}}.
\newblock Cambridge University Press,
2014.
\newblock

\bibitem{Kaminski:2009fm}
W.~Kaminski, M.~Kisielowski, and J.~Lewandowski, ``{Spin-Foams for All Loop
  Quantum Gravity},'' {\em Class. Quant. Grav.} {\bf 27} (2010) 095006,
  \href{http://arXiv.org/abs/0909.0939}{{\tt arXiv:0909.0939}}. [Erratum:
  Class.Quant.Grav. 29, 049502 (2012)].

\bibitem{Dittrich:2018xuk}
B.~Dittrich, C.~Goeller, E.~R. Livine, and A.~Riello, ``{Quasi-local
  holographic dualities in non-perturbative 3d quantum gravity},'' {\em Class.
  Quant. Grav.} {\bf 35} (2018), no.~13, 13LT01,
\href{http://arXiv.org/abs/1803.02759}{{\tt arXiv:1803.02759}}.

\bibitem{Dittrich:2014ala}
B.~Dittrich, ``{The continuum limit of loop quantum gravity - a framework for
  solving the theory},'' in {\em Loop Quantum Gravity, The First Thirty Years},
  A.~Abhay and J.~Pullin, eds., vol.~4.
\newblock World Scientific, 2017.
\newblock
\href{http://arXiv.org/abs/1409.1450}{{\tt arXiv:1409.1450}}.
\newblock

\bibitem{Asante:2020qpa}
S.~K. Asante, B.~Dittrich, and H.~M. Haggard, ``{Effective Spin Foam Models for
  Four-Dimensional Quantum Gravity},'' {\em Phys. Rev. Lett.} {\bf 125} (2020),
  no.~23, 231301, \href{http://arXiv.org/abs/2004.07013}{{\tt
  arXiv:2004.07013}}.

\bibitem{Ashtekar:aa}
A.~Ashtekar, C.~Beetle, and S.~Fairhurst, ``Isolated Horizons: A Generalization
  of Black Hole Mechanics,'' {\em Class. Quant. Grav.} {\bf 16} (1999),
  no.~L1--L7, \href{http://arXiv.org/abs/gr-qc/9812065}{{\tt
  arXiv:gr-qc/9812065}}.

\bibitem{Ashtekar:2004aa}
A.~Ashtekar and B.~Krishnan, ``Isolated and Dynamical Horizons and Their
  Applications,'' {\em Living Reviews in Relativity} {\bf 7} (2004), no.~1, 10.

\bibitem{Ashtekar:2001is}
A.~Ashtekar, C.~Beetle, and J.~Lewandowski, ``{Mechanics of rotating isolated
  horizons},'' {\em Phys. Rev. D} {\bf 64} (2001) 044016,
\href{http://arXiv.org/abs/gr-qc/0103026}{{\tt arXiv:gr-qc/0103026}}.

\bibitem{Lewandowski:2002ua}
J.~Lewandowski and T.~Pawlowski, ``{Extremal isolated horizons: A Local
  uniqueness theorem},'' {\em Class. Quant. Grav.} {\bf 20} (2003) 587--606,
  \href{http://arXiv.org/abs/gr-qc/0208032}{{\tt arXiv:gr-qc/0208032}}.

\bibitem{Lewandowski:1999zs}
J.~Lewandowski, ``{Space-times admitting isolated horizons},'' {\em Class.
  Quant. Grav.} {\bf 17} (2000) L53--L59,
  \href{http://arXiv.org/abs/gr-qc/9907058}{{\tt arXiv:gr-qc/9907058}}.

\bibitem{Ashtekar:2021wld}
A.~Ashtekar, N.~Khera, M.~Kolanowski, and J.~Lewandowski, ``{Non-expanding
  horizons: multipoles and the symmetry group},'' {\em JHEP} {\bf 01} (2022)
  028, \href{http://arXiv.org/abs/2111.07873}{{\tt arXiv:2111.07873}}.

\bibitem{Ashtekar:2024stm}
A.~Ashtekar and S.~Speziale, ``{Null infinity and horizons: A new approach to
  fluxes and charges},'' {\em Phys. Rev. D} {\bf 110} (2024), no.~4, 044049,
  \href{http://arXiv.org/abs/2407.03254}{{\tt arXiv:2407.03254}}.

\bibitem{Ashtekar:2021kqj}
A.~Ashtekar, N.~Khera, M.~Kolanowski, and J.~Lewandowski, ``{Charges and fluxes
  on (perturbed) non-expanding horizons},'' {\em JHEP} {\bf 02} (2022) 066,
  \href{http://arXiv.org/abs/2112.05608}{{\tt arXiv:2112.05608}}.

\bibitem{Rovelli:1990pi}
C.~Rovelli, ``{Quantum Reference Frames},'' {\em Class. Quant. Grav.} {\bf 8}
  (1991) 317--332.

\bibitem{Giacomini:2019aa}
F.~Giacomini, E.~Castro-Ruiz, and {\v C}.~Brukner, ``Quantum mechanics and the
  covariance of physical laws in quantum reference frames,'' {\em Nature
  Communications} {\bf 10} (2019), no.~1, 494.

\bibitem{Hardy:2019cef}
L.~Hardy, ``{Implementation of the Quantum Equivalence Principle},'' in {\em
  {Progress and Visions in Quantum Theory in View of Gravity}: {Bridging
  foundations of physics and mathematics}}.
\newblock 3, 2019.
\newblock \href{http://arXiv.org/abs/1903.01289}{{\tt arXiv:1903.01289}}.

\bibitem{Vanrietvelde:2018pgb}
A.~Vanrietvelde, P.~A. Hoehn, F.~Giacomini, and E.~Castro-Ruiz, ``{A change of
  perspective: switching quantum reference frames via a perspective-neutral
  framework},'' {\em Quantum} {\bf 4} (2020) 225,
  \href{http://arXiv.org/abs/1809.00556}{{\tt arXiv:1809.00556}}.

\bibitem{Hoehn:2019fsy}
P.~A. Hoehn, A.~R.~H. Smith, and M.~P.~E. Lock, ``{Trinity of relational
  quantum dynamics},'' {\em Phys. Rev. D} {\bf 104} (2021), no.~6, 066001,
  \href{http://arXiv.org/abs/1912.00033}{{\tt arXiv:1912.00033}}.

\bibitem{delaHamette:2020dyi}
A.-C. de~la Hamette and T.~D. Galley, ``{Quantum reference frames for general
  symmetry groups},'' {\em Quantum} {\bf 4} (2020) 367,
  \href{http://arXiv.org/abs/2004.14292}{{\tt arXiv:2004.14292}}.

\bibitem{Loveridge2018}
L.~Loveridge, T.~Miyadera, and P.~Busch, ``Symmetry, Reference Frames, and
  Relational Quantities in Quantum Mechanics,'' {\em Foundations of Physics}
  {\bf 48} (2018), no.~2, 135--198.

\bibitem{Balachandran:1994up}
A.~P. Balachandran, L.~Chandar, and A.~Momen, ``{Edge states in gravity and
  black hole physics},'' {\em Nucl. Phys. B} {\bf 461} (1996) 581--596,
\href{http://arXiv.org/abs/gr-qc/9412019}{{\tt arXiv:gr-qc/9412019}}.

\bibitem{Carlip:1996yb}
S.~Carlip, ``{The Statistical mechanics of the three-dimensional Euclidean
  black hole},'' {\em Phys. Rev. D} {\bf 55} (1997) 878--882,
\href{http://arXiv.org/abs/gr-qc/9606043}{{\tt arXiv:gr-qc/9606043}}.

\bibitem{Freidel:2015gpa}
L.~Freidel and A.~Perez, ``{Quantum gravity at the corner},'' {\em Universe}
  {\bf 4} (2018), no.~10, 107, \href{http://arXiv.org/abs/1507.02573}{{\tt
  arXiv:1507.02573}}.

\bibitem{Wieland:2017zkf}
W.~Wieland, ``{New boundary variables for classical and quantum gravity on a
  null surface},'' {\em Class. Quantum Grav.} {\bf 34} (2017) 215008,
\href{http://arXiv.org/abs/1704.07391}{{\tt arXiv:1704.07391}}.

\bibitem{Wieland:2017cmf}
W.~Wieland, ``{Fock representation of gravitational boundary modes and the
  discreteness of the area spectrum},'' {\em Ann. Henri Poincar{\'e}} {\bf 18}
  (2017) 3695--3717,
\href{http://arXiv.org/abs/1706.00479}{{\tt arXiv:1706.00479}}.

\bibitem{Wieland:2021vef}
W.~Wieland, ``{Gravitational SL(2, \ensuremath{\mathbb{R}}) algebra on the
  light cone},'' {\em JHEP} {\bf 07} (2021) 057,
  \href{http://arXiv.org/abs/2104.05803}{{\tt arXiv:2104.05803}}.

\bibitem{Speranza:2017gxd}
A.~J. Speranza, ``{Local phase space and edge modes for
  diffeomorphism-invariant theories},'' {\em JHEP} {\bf 02} (2018) 021,
  \href{http://arXiv.org/abs/1706.05061}{{\tt arXiv:1706.05061}}.

\bibitem{Freidel:2021cjp}
L.~Freidel, R.~Oliveri, D.~Pranzetti, and S.~Speziale, ``{Extended corner
  symmetry, charge bracket and Einstein\textquoteright{}s equations},'' {\em
  JHEP} {\bf 09} (2021) 083, \href{http://arXiv.org/abs/2104.12881}{{\tt
  arXiv:2104.12881}}.

\bibitem{Freidel:2021fxf}
L.~Freidel, R.~Oliveri, D.~Pranzetti, and S.~Speziale, ``{The Weyl BMS group
  and Einstein{\textquoteright}s equations},'' {\em JHEP} {\bf 07} (2021) 170,
  \href{http://arXiv.org/abs/2104.05793}{{\tt arXiv:2104.05793}}.

\bibitem{Freidel:2020xyx}
L.~Freidel, M.~Geiller, and D.~Pranzetti, ``{Edge modes of gravity. Part I.
  Corner potentials and charges},'' {\em JHEP} {\bf 11} (2020) 026,
  \href{http://arXiv.org/abs/2006.12527}{{\tt arXiv:2006.12527}}.

\bibitem{Carrozza:2021gju}
S.~Carrozza and P.~A. Hoehn, ``{Edge modes as reference frames and boundary
  actions from post-selection},'' {\em JHEP} {\bf 02} (2022) 172,
  \href{http://arXiv.org/abs/2109.06184}{{\tt arXiv:2109.06184}}.

\bibitem{Kabel:2023jve}
V.~Kabel, {\v{C}}.~Brukner, and W.~Wieland, ``{Quantum reference frames at the
  boundary of spacetime},'' {\em Phys. Rev. D} {\bf 108} (2023), no.~10,
  106022, \href{http://arXiv.org/abs/2302.11629}{{\tt arXiv:2302.11629}}.

\bibitem{Freidel:2023bnj}
L.~Freidel, M.~Geiller, and W.~Wieland, ``{Corner symmetry and quantum
  geometry},'' in {\em Handbook of Quantum Gravity}, L.~M. Cosimo~Bambi and
  I.~Shapiro, eds.
\newblock Springer, 2023.
\newblock \href{http://arXiv.org/abs/2302.12799}{{\tt arXiv:2302.12799}}.

\bibitem{Giesel:2024xtb}
K.~Giesel, V.~Kabel, and W.~Wieland, ``{Linking edge modes and geometrical
  clocks in linearized gravity},'' {\em Phys. Rev. D} {\bf 112} (2025), no.~6,
  064063, \href{http://arXiv.org/abs/2410.17339}{{\tt arXiv:2410.17339}}.

\bibitem{Fewster:2024pur}
J.~C. Fewster, D.~W. Janssen, L.~D. Loveridge, K.~Rejzner, and J.~Waldron,
  ``{Quantum Reference Frames, Measurement Schemes and the Type of Local
  Algebras in Quantum Field Theory},'' {\em Commun. Math. Phys.} {\bf 406}
  (2025), no.~1, 19, \href{http://arXiv.org/abs/2403.11973}{{\tt
  arXiv:2403.11973}}.

\bibitem{Fewster:2025ijg}
C.~J. Fewster, D.~W. Janssen, and K.~Rejzner, ``{Semi-local observables, edge
  modes and quantum reference frames in quantum electromagnetism: an algebraic
  approach},'' \href{http://arXiv.org/abs/2508.20939}{{\tt arXiv:2508.20939}}.

\bibitem{Giddings:2019hjc}
S.~B. Giddings, ``{Gravitational dressing, soft charges, and perturbative
  gravitational splitting},'' {\em Phys. Rev. D} {\bf 100} (2019), no.~12,
  126001, \href{http://arXiv.org/abs/1903.06160}{{\tt arXiv:1903.06160}}.

\bibitem{Donnelly:2016auv}
W.~Donnelly and L.~Freidel, ``{Local subsystems in gauge theory and gravity},''
  {\em JHEP} {\bf 09} (2016) 102, \href{http://arXiv.org/abs/1601.04744}{{\tt
  arXiv:1601.04744}}.

\bibitem{Donnelly:2016rvo}
W.~Donnelly and S.~B. Giddings, ``{Observables, gravitational dressing, and
  obstructions to locality and subsystems},'' {\em Phys. Rev. D} {\bf 94}
  (2016), no.~10, 104038,
\href{http://arXiv.org/abs/1607.01025}{{\tt arXiv:1607.01025}}.

\bibitem{Donnelly:2015hta}
W.~Donnelly and S.~B. Giddings, ``{Diffeomorphism-invariant observables and
  their nonlocal algebra},'' {\em Phys. Rev. D} {\bf 93} (2016), no.~2, 024030,
  \href{http://arXiv.org/abs/1507.07921}{{\tt arXiv:1507.07921}}. [Erratum:
  Phys.Rev.D 94, 029903 (2016)].

\bibitem{Donnelly:2017jcd}
W.~Donnelly and S.~B. Giddings, ``{How is quantum information localized in
  gravity?},'' {\em Phys. Rev. D} {\bf 96} (2017), no.~8, 086013,
  \href{http://arXiv.org/abs/1706.03104}{{\tt arXiv:1706.03104}}.

\bibitem{AshtekarNullInfinity}
A.~Ashtekar, {\em {Asymptotic Quantization}}.
\newblock Bibliopolis, Napoli, 1987.
\newblock Based on 1984 Naples Lectures.

\bibitem{Barnich:2011mi}
G.~Barnich and C.~Troessaert, ``{BMS charge algebra},'' {\em JHEP} {\bf 12}
  (2011) 105,
\href{http://arXiv.org/abs/1106.0213}{{\tt arXiv:1106.0213}}.

\bibitem{Strominger:2017zoo}
A.~Strominger, {\em {Lectures on the Infrared Structure of Gravity and Gauge
  Theory}}.
\newblock Princeton University Press, Princeton, 2018.
\newblock
\href{http://arXiv.org/abs/1703.05448}{{\tt arXiv:1703.05448}}.
\newblock

\bibitem{Pasterski:2015zua}
S.~Pasterski, ``{Asymptotic Symmetries and Electromagnetic Memory},'' {\em
  JHEP} {\bf 09} (2017) 154, \href{http://arXiv.org/abs/1505.00716}{{\tt
  arXiv:1505.00716}}.

\bibitem{Pasterski:2021raf}
S.~Pasterski, M.~Pate, and A.-M. Raclariu, ``{Celestial Holography},'' in {\em
  {Snowmass 2021}}.
\newblock 11, 2021.
\newblock \href{http://arXiv.org/abs/2111.11392}{{\tt arXiv:2111.11392}}.

\bibitem{Wieland:2020gno}
W.~Wieland, ``{Null infinity as an open Hamiltonian system},'' {\em JHEP} {\bf
  04} (2021) 095, \href{http://arXiv.org/abs/2012.01889}{{\tt
  arXiv:2012.01889}}.

\bibitem{Freidel:2021dfs}
L.~Freidel, D.~Pranzetti, and A.-M. Raclariu, ``{Sub-subleading soft graviton
  theorem from asymptotic Einstein{\textquoteright}s equations},'' {\em JHEP}
  {\bf 05} (2022) 186, \href{http://arXiv.org/abs/2111.15607}{{\tt
  arXiv:2111.15607}}.

\bibitem{Ruzziconi:2025fuy}
R.~Ruzziconi and C.~Zwikel, ``{Celestial $Lw_{1+\infty}$ Symmetries and
  Subleading Phase Space of Null Hypersurfaces},''
  \href{http://arXiv.org/abs/2511.07525}{{\tt arXiv:2511.07525}}.

\bibitem{Wittenreal}
E.~Witten, ``{Quantum field theory and the Jones polynomial},'' {\em {Comm.
  Math. Phys.}} {\bf 121} (1989) 351--399.

\bibitem{ChernSimonsBook}
M.~Mari{\~n}o, {\em Chern-Simons Theory, Matrix Models, and Topological
  Strings}.
\newblock Oxford University Press, 09, 2005.

\bibitem{1987CrnkovicWitten}
C.~{Crnkovic} and E.~{Witten}, ``{Covariant description of canonical formalism
  in geometrical theories.},'' in {\em Three Hundred Years of Gravitation},
  S.~W. {Hawking} and W.~{Israel}, eds., pp.~676--684.
\newblock Cambridge University Pressbridge, 1987.

\bibitem{Ashtekar:1987hia}
A.~Ashtekar, L.~Bombelli, and R.~Koul, ``{Phase space formulation of general
  relativity without a 3+1 splitting},'' {\em Lect. Notes Phys.} {\bf 278}
  (1987) 356--359.

\bibitem{Wald:1999wa}
R.~M. Wald and A.~Zoupas, ``{A General definition of `conserved quantities' in
  general relativity and other theories of gravity},'' {\em Phys. Rev. D} {\bf
  61} (2000) 084027,
\href{http://arXiv.org/abs/gr-qc/9911095}{{\tt arXiv:gr-qc/9911095}}.

\bibitem{Ashtekar:1991my}
A.~Ashtekar and C.~Rovelli, ``{A Loop representation for the quantum Maxwell
  field},'' {\em Class. Quant. Grav.} {\bf 9} (1992) 1121--1150,
  \href{http://arXiv.org/abs/hep-th/9202063}{{\tt arXiv:hep-th/9202063}}.

\bibitem{Ashtekar:1991mz}
A.~Ashtekar, C.~Rovelli, and L.~Smolin, ``{Gravitons and loops},'' {\em Phys.
  Rev. D} {\bf 44} (1991) 1740--1755,
  \href{http://arXiv.org/abs/hep-th/9202054}{{\tt arXiv:hep-th/9202054}}.

\bibitem{Varadarajan:1999it}
M.~Varadarajan, ``{Fock representations from U(1) holonomy algebras},'' {\em
  Phys. Rev. D} {\bf 61} (2000) 104001,
  \href{http://arXiv.org/abs/gr-qc/0001050}{{\tt arXiv:gr-qc/0001050}}.

\bibitem{newmanpenrose}
E.~Newman and R.~Penrose, ``An Approach to Gravitational Radiation by a Method
  of Spin Coefficients,'' {\em Journal of Mathematical Physics} {\bf 3} (1962),
  no.~3, 566--578.

\bibitem{penroserindler}
R.~Penrose and W.~Rindler, {\em Spinors and Space-Time, Two-Spinor Calculus and
  Relativistic Fields}, vol.~1 and 2.
\newblock Cambridge University Press, Cambridge, 1984.

\bibitem{PhysRevD.11.395}
J.~Kogut and L.~Susskind, ``Hamiltonian formulation of Wilson's lattice gauge
  theories,'' {\em Phys. Rev. D} {\bf 11} (Jan, 1975) 395--408.

\bibitem{PhysRevD.19.619}
S.~D. Drell, H.~R. Quinn, B.~Svetitsky, and M.~Weinstein, ``Quantum
  electrodynamics on a lattice: A Hamiltonian variational approach to the
  physics of the weak-coupling region,'' {\em Phys. Rev. D} {\bf 19} (Jan,
  1979) 619--638.

\bibitem{Bojowald:1999fw}
M.~Bojowald, ``{Abelian BF theory and spherically symmetric
  electromagnetism},'' {\em J. Math. Phys.} {\bf 41} (2000) 4313--4329,
  \href{http://arXiv.org/abs/hep-th/9908170}{{\tt arXiv:hep-th/9908170}}.

\bibitem{Drobinski:2017kfm}
P.~Drobi{\'n}ski and J.~Lewandowski, ``{Continuum approach to the BF vacuum:
  The U(1) case},'' {\em Phys. Rev. D} {\bf 96} (2017), no.~12, 126011,
  \href{http://arXiv.org/abs/1705.09836}{{\tt arXiv:1705.09836}}.

\bibitem{Magnifico:2020bqt}
G.~Magnifico, T.~Felser, P.~Silvi, and S.~Montangero, ``{Lattice quantum
  electrodynamics in (3+1)-dimensions at finite density with tensor
  networks},'' {\em Nature Commun.} {\bf 12} (2021), no.~1, 3600,
  \href{http://arXiv.org/abs/2011.10658}{{\tt arXiv:2011.10658}}.

\bibitem{Smolin:1992wj}
L.~Smolin, ``{The $G_{\text{Newton}}$ to 0 limit of Euclidean quantum
  gravity},'' {\em Class. Quant. Grav.} {\bf 9} (1992) 883--894,
  \href{http://arXiv.org/abs/hep-th/9202076}{{\tt arXiv:hep-th/9202076}}.

\bibitem{Varadarajan:2018tei}
M.~Varadarajan, ``{Constraint algebra in Smolins' $G\rightarrow 0$ limit of 4d
  Euclidean gravity},'' {\em Phys. Rev. D} {\bf 97} (2018), no.~10, 106007,
  \href{http://arXiv.org/abs/1802.07033}{{\tt arXiv:1802.07033}}.

\bibitem{Zarate:2025erv}
M.~R. Zarate and T.~Thiemann, ``{Hamiltonian renormalisation IX. $U(1)^3$
  quantum gravity},'' \href{http://arXiv.org/abs/2505.13037}{{\tt
  arXiv:2505.13037}}.

\bibitem{Bakhoda:2024mth}
S.~Bakhoda and Y.~Ma, ``{Geometrical quantum time in the U(1)$^{3}$ model of
  Euclidean quantum gravity},'' {\em Commun. Theor. Phys.} {\bf 77} (2025),
  no.~5, 055401, \href{http://arXiv.org/abs/2411.19435}{{\tt
  arXiv:2411.19435}}.

\bibitem{Bakhoda:2020ril}
S.~Bakhoda and T.~Thiemann, ``{Reduced phase space approach to the $U(1)^3$
  model for Euclidean quantum gravity},'' {\em Class. Quant. Grav.} {\bf 38}
  (2021), no.~21, 215006, \href{http://arXiv.org/abs/2010.16351}{{\tt
  arXiv:2010.16351}}.

\bibitem{Bakhoda:2020fiy}
S.~Bakhoda and T.~Thiemann, ``{Covariant origin of the U(1)$^{3}$ model for
  Euclidean quantum gravity},'' {\em Class. Quant. Grav.} {\bf 39} (2022),
  no.~2, 025006, \href{http://arXiv.org/abs/2011.00031}{{\tt
  arXiv:2011.00031}}.

\bibitem{Thiemann:2022all}
T.~Thiemann, ``{Exact quantisation of U(1)$^{3}$ quantum gravity via
  exponentiation of the hypersurface deformation algebroid},'' {\em Class.
  Quant. Grav.} {\bf 40} (2023), no.~24, 245003,
  \href{http://arXiv.org/abs/2207.08302}{{\tt arXiv:2207.08302}}.

\bibitem{Sahlmann:2024pba}
H.~Sahlmann and W.~Sherif, ``{Towards quantum gravity with neural networks:
  solving the quantum Hamilton constraint of U(1) BF theory},'' {\em Class.
  Quant. Grav.} {\bf 41} (2024), no.~22, 225014,
  \href{http://arXiv.org/abs/2402.10622}{{\tt arXiv:2402.10622}}.

\bibitem{wald}
R.~M. Wald, {\em General Relativity}.
\newblock The University of Chicago Press, Chicago, London, 1984.

\bibitem{Riello:2022din}
A.~Riello and M.~Schiavina, ``{Hamiltonian gauge theory with corners:
  constraint reduction and flux superselection},''
  \href{http://arXiv.org/abs/2207.00568}{{\tt arXiv:2207.00568}}.

\bibitem{Blohmann:2022yqo}
C.~Blohmann, M.~Schiavina, and A.~Weinstein, ``{A Lie{\textendash}Rinehart
  algebra in general relativity},'' {\em Pure Appl. Math. Quart.} {\bf 19}
  (2023), no.~4, 1733--1777, \href{http://arXiv.org/abs/2201.02883}{{\tt
  arXiv:2201.02883}}.

\bibitem{Ashtekar:1993wf}
A.~Ashtekar and J.~Lewandowski, ``{Representation theory of analytic holonomy
  C* algebras},'' in {\em Knots and Quantum Gravity}, J.Baez, ed.
\newblock Oxford University Press, 1993.
\newblock
\href{http://arXiv.org/abs/gr-qc/9311010}{{\tt arXiv:gr-qc/9311010}}.
\newblock

\bibitem{Ashtekar:1995zh}
A.~Ashtekar, J.~Lewandowski, D.~Marolf, J.~Mourao, and T.~Thiemann,
  ``{Quantization of diffeomorphism invariant theories of connections with
  local degrees of freedom},'' {\em J. Math. Phys.} {\bf 36} (1995) 6456--6493,
  \href{http://arXiv.org/abs/gr-qc/9504018}{{\tt arXiv:gr-qc/9504018}}.

\bibitem{Ashtekar:1994mh}
A.~Ashtekar and J.~Lewandowski, ``{Projective techniques and functional
  integration for gauge theories},'' {\em J. Math. Phys.} {\bf 36} (1995)
  2170--2191,
\href{http://arXiv.org/abs/gr-qc/9411046}{{\tt arXiv:gr-qc/9411046}}.

\bibitem{Fleischhack:2004jc}
C.~Fleischhack, ``{Representations of the Weyl algebra in quantum geometry},''
  {\em Commun. Math. Phys.} {\bf 285} (2009) 67--140,
  \href{http://arXiv.org/abs/math-ph/0407006}{{\tt arXiv:math-ph/0407006}}.

\bibitem{ALvacuum}
J.~Lewandowski, A.~Okolow, H.~Sahlmann, and T.~Thiemann, ``{Uniqueness of
  diffeomorphism invariant states on holonomy-flux algebras},'' {\em Commun.
  Math. Phys.} {\bf 267} (2006) 703--733,
\href{http://arXiv.org/abs/gr-qc/0504147}{{\tt arXiv:gr-qc/0504147}}.

\bibitem{Asante2023}
S.~K. Asante, B.~Dittrich, and S.~Steinhaus, ``Spin Foams, Refinement Limit,
  and Renormalization,'' in {\em Handbook of Quantum Gravity}, C.~Bambi,
  L.~Modesto, and I.~Shapiro, eds., pp.~1--37.
\newblock Springer Nature Singapore, Singapore, 2023.
\newblock \href{http://arXiv.org/abs/2211.09578}{{\tt arXiv:2211.09578}}.

\bibitem{Freidel:2020svx}
L.~Freidel, M.~Geiller, and D.~Pranzetti, ``{Edge modes of gravity. Part II.
  Corner metric and Lorentz charges},'' {\em JHEP} {\bf 11} (2020) 027,
  \href{http://arXiv.org/abs/2007.03563}{{\tt arXiv:2007.03563}}.

\bibitem{Freidel:2020ayo}
L.~Freidel, M.~Geiller, and D.~Pranzetti, ``{Edge modes of gravity. Part III.
  Corner simplicity constraints},'' {\em JHEP} {\bf 01} (2021) 100,
  \href{http://arXiv.org/abs/2007.12635}{{\tt arXiv:2007.12635}}.

\bibitem{Gomes:2016mwl}
H.~Gomes and A.~Riello, ``{The observer's ghost: notes on a field space
  connection},'' {\em JHEP} {\bf 05} (2017) 017,
\href{http://arXiv.org/abs/1608.08226}{{\tt arXiv:1608.08226}}.

\bibitem{Assanioussi:2023jyq}
M.~Assanioussi, J.~Kowalski-Glikman, I.~M{\"a}kinen, and L.~Varrin, ``{On the
  covariant formulation of gauge theories with boundaries},'' {\em Class.
  Quant. Grav.} {\bf 41} (2024), no.~11, 115007,
  \href{http://arXiv.org/abs/2312.01918}{{\tt arXiv:2312.01918}}.

\bibitem{Langenscheidt:2024nyw}
S.~Langenscheidt and D.~Oriti, ``{New edge modes and corner charges for
  first-order symmetries of 4D gravity},'' {\em Class. Quant. Grav.} {\bf 42}
  (2025), no.~7, 075010, \href{http://arXiv.org/abs/2408.01809}{{\tt
  arXiv:2408.01809}}.

\bibitem{Neri:2025fsh}
G.~Neri and L.~Varrin, ``{Orbit method for quantum corner symmetries},'' {\em
  Phys. Rev. D} {\bf 112} (2025), no.~10, 104010,
  \href{http://arXiv.org/abs/2507.10683}{{\tt arXiv:2507.10683}}.

\bibitem{Wieland:2021eth}
W.~Wieland, ``{Barnich\textendash{}Troessaert bracket as a Dirac bracket on the
  covariant phase space},'' {\em Class. Quant. Grav.} {\bf 39} (2022), no.~2,
  025016, \href{http://arXiv.org/abs/2104.08377}{{\tt arXiv:2104.08377}}.

\bibitem{Wieland:2024dop}
W.~Wieland, ``{Quantum geometry of the null cone},'' {\em Phys. Rev. D} {\bf
  112} (2025), no.~4, 044042, \href{http://arXiv.org/abs/2401.17491}{{\tt
  arXiv:2401.17491}}.

\bibitem{Wieland:2025qgx}
W.~Wieland, ``{Quantum geometry of the light cone: Fock representation and
  spectrum of radiated power},'' {\em Class. Quant. Grav.} {\bf 42} (2025),
  no.~19, 195006, \href{http://arXiv.org/abs/2504.10802}{{\tt
  arXiv:2504.10802}}.

\bibitem{shadowstats}
A.~Ashtekar, S.~Fairhurst, and J.~L. Willis, ``{Quantum gravity, shadow states,
  and quantum mechanics},'' {\em Class. Quantum Grav.} {\bf 20} (March, 2003)
  1031--1061, \href{http://arXiv.org/abs/gr-qc/0207106v3}{{\tt
  arXiv:gr-qc/0207106v3}}.

\bibitem{Ashtekar:2001xp}
A.~Ashtekar and J.~Lewandowski, ``{Relation between polymer and Fock
  excitations},'' {\em Class. Quant. Grav.} {\bf 18} (2001) L117--L128,
  \href{http://arXiv.org/abs/gr-qc/0107043}{{\tt arXiv:gr-qc/0107043}}.

\bibitem{lqcmath}
A.~Ashtekar, M.~Bojowald, and J.~Lewandowski, ``{Mathematical structure of loop
  quantum cosmology},'' {\em Advances in Theoretical and Mathematical Physics}
  {\bf 7} (2003), no.~1, 233--268,
  \href{http://arXiv.org/abs/gr-qc/0304074v4}{{\tt arXiv:gr-qc/0304074v4}}.

\bibitem{Majid_1995}
S.~Majid, {\em Foundations of Quantum Group Theory}.
\newblock Cambridge University Press, 1995.

\bibitem{Ciambelli:2023mir}
L.~Ciambelli, L.~Freidel, and R.~G. Leigh, ``{Null Raychaudhuri: Canonical
  Structure and the Dressing Time},''
  \href{http://arXiv.org/abs/2309.03932}{{\tt arXiv:2309.03932}}.

\bibitem{CFTbook}
D.~S. Philippe Di~Francecso, Pierre~Mathieu, {\em Conformal Field Theory}.
\newblock Springer, 1997.

\end{thebibliography}
\end{document}